\documentclass[12pt]{iopart}
\usepackage{stackrel}
\usepackage{iopams}
\usepackage{graphicx}
\usepackage{color}

\usepackage{ulem}
\usepackage{dsfont}
\usepackage{bm}
\usepackage[utf8]{inputenc}
\usepackage[english]{babel}
\usepackage{feynmp}
\usepackage{tikz}
\usetikzlibrary{shapes,arrows,positioning,automata,backgrounds,calc,er,patterns}
\usepackage{hyperref}

\usepackage{iopams}  
\usepackage{pst-pdf}

\begin{document}




\title{Single file diffusion meets Feynman path integral}

\author{Pavel Castro-Villarreal$^{a}$, Claudio Contreras-Aburto\footnote{Present address: Facultad de Física, Universidad Veracruzana, Circuito Aguirre Beltrán S/N, 91000 Xalapa, Veracruz, M\'exico.},  Sendic Estrada-Jim\'enez, Idrish Huet-Hern\'andez$^{b}$, Oscar V\'azquez-Rodr\'iguez }%
\address{Facultad de Ciencias en F\'isica y Matem\'aticas, Universidad Aut\'onoma de Chiapas, Carretera Emiliano Zapata, Km. 8, Rancho San Francisco, C. P. 29050, Tuxtla Guti\'errez, Chiapas, M\'exico.}

\ead{$^{a}$pcastrov@unach.mx}
\ead{$^{b}$idrish.huet@unach.mx}
\vspace{10pt}
\begin{indented}
\item[]May 2021
\end{indented}


\begin{abstract} 
The path-integral representation of Smoluchowski equation  is exploited to explore the stochastic dynamics of a tagged Brownian particle within an interacting system where hydrodynamic effects are neglected. In particular, this formalism is applied to a particle system confined to a one-dimensional infinite line aiming to investigate the single-file diffusion phenomenon in this scenario. In particular, the path-integral method is contrasted against the standard many-particle Langevin equation for a system of interacting Brownian particles in a harmonic chain model, exhibiting excellent agreement; in this case of study a formula defined on the whole time-scale for the mean-square displacement, in the thermodynamic limit, is found for the tracer particle in terms of Bessel functions, recovering also the single-file regime. Additionally, a Brownian particle system  with paramagnetic interactions is considered near crystallization where the total interaction potential is roughly a harmonic potential. Taking advantage of the path-integral formalism a simple perturbation treatment is carried out to investigate the single file diffusion behavior when temperature is increased away from the crystal phase.  \end{abstract}

\section{Introduction}
Single file diffusion (SFD) is a term coined to describe the  diffusion of particles constrained to a narrow channel with the condition that the mutual passage of particles is excluded. Ranging from the original evidence of the reduced transport of ion potassium through a giant nerve axon membrane of {\it Sepia Officinalis} \cite{HODGKIN1955} to molecular diffusion in zeolites \cite{Karger2003} there has been an intense and continued research activity in order to understand the anomalous character of SFD. In particular, during past decades there has been a plethora of experiments and theoretical analysis to provide confirmation of the universal validity of its subdiffusion behavior, that is,  the mean-square displacement (MSD) of a tagged particle at time $t$ turns out to be proportional to $t^{1/2}$.  Let us summarize the developments most significant for the present work.  On the experimental side,  by confining paramagnetic colloids to a circular channel made by photolitography  \cite{Bechinger2000},  and then   creating one-dimensional channels with scanned optical tweezers to avoid hydrodynamic effects \cite{Lutz2004, LutzPRL2004} it was possible  to show a non-Fickian long time behavior in an anomalous SF process. In contrast, the  diffusion of millimetric balls mimicking vortices in a superconductor in a circular channel showed a significant deviation from the usual SF \cite{Coupier-2006-1}; moreover, using a fluctuating modulated potential a significant enhancement of the tagged particle's diffusion contrasted with a bald channel was reported in \cite{Coupier_2007}. 
However, the transport of macroscopic charged beads interacting with an screened electrostatic potential behaves as the SFD \cite{coste-2010}. On the theoretical side, the anomalous SF diffusion was set on a firm footing in \cite{Kollmann2003} using a general theory based on the generalized Smoluchowski equation \cite{NAGELE1996215}, where the proportionality pre-factor in the MSD was found to be given in terms of the compressibility of the system. Also, an exact analytical expression for the tagged particle's probability density function using a Bethe ansatz was derived in \cite{Lizana2008}. Introducing an harmonization techinque to map a many-body interacting system into a system of beads interconnected by harmonic springs it was possible to derive a stochastic fractional Langevin equation for the tracer particle that captures the essence of the SF behavior \cite{Lizana2010}.

A Brownian harmonic chain was considered and treated by means of the Langevin equation in \cite{Delfau2011} where different size-effects of the single-file diffusion were found, in particular, for a finite number of particles it was found that after a sufficiently long time the system presents free diffusion with a reduced diffusion coefficient $D_{0}/N$, where $D_{0}$ is the free particle self-diffusion coefficient. The SFD behavior was corroborated for point-like particles in a file with a periodic substrate \cite{Taloni-Marcheroni}, and a rich variety of absorbate phases were found for charged and superparamagnetic colloidal particles in a similar file using numerical simulations \cite{Chava-Ramon-2007,Chava-Ramon-2008}.  A comprehensive and profound  revision of the SFD with a scientific and technological perspective can be found in \cite{Taloni-Ramon2017}.

According to these results the transport phenomenon behind single file diffusion is the result emerging from the collective interactions of Brownian particles confined to a straight line with the constraint that particles cannot pass each other through. Even though the SF behavior $\sim t^{1/2}$ is found in all the aforementioned physical scenarios it is natural to ask whether the single file behavior is the unique file diffusion occurring in the long time regime. For instance, what would have happened with SF behavior in the low, or high-temperature regime in an open long line?, or how is the SF affected by the thermodynamic phase of the system? Even if the general answer to these questions lies beyond the scope of this work, here we investigated the diffusive regime that emerges in the file's dynamics when the temperature is increased in a file system at the crystalline phase. Although Brownian dynamics has already been approached using different theoretical frameworks such as many-particle Langevin equations, generalized Smoluchowski equation or Brownian dynamics simulations adapted to various physical scenarios \cite{Dhont}, in this work we have found an useful approach through the path-integral formalism \cite{Brush61,WIEGEL197557,Zinn-JustinBook, Chaichian}. In this paper we approach Brownian dynamics (BD) from the perspective of Feynman's path integral formalism.  As a starting point we consider Brownian dynamics based on the so called generalized Smoluchowski diffusion equation \cite{NAGELE1996215} (here simply called  Smoluchowski equation), the basic treatment consists then of carrying out 
simple transformations that allow to convert the Smoluchowski equation into a high-dimensional Schr$\ddot{\rm o}$dinger type equation \cite{Olindo2009} at which point a standard treatment \cite{Zinn-JustinBook, Chaichian} can be followed using the path integral formalism. Using this idea we study a fluid with strong interactions near the crystallization state, first through  the Frenkel-Kontoreva model with a flat substrate, and second for an approximation of the paramagnetic interaction. 

It is shown that the results coming from the path integral approach, and those from the Langevin description,  analytical and numerical,  are identical. This fact places the BD path-integral formulation on a firm ground with an outlook to perform further more complex calculations. For instance, in \cite{Olindo2009}, this formalism was used to improve a version of the Monte-Carlo algorithm in order to investigate the long-time dynamics of several molecular systems. Here we take advantage of the BD path-integral formulation to study analytically a simple temperature perturbation of the crystalline state of the system. Our findings show a breakdown of the typical behaviour of the SFD after a sufficiently long time or after a certain temperature. From the physical point of view this result is reasonable and should be expected since, as the thermal agitation increases in an open one dimensional system, the particle mobility grows, and thus the particle diffusion is increased. This phenomenon, observed at the level of the MSD, constitutes evidence that this effect is easier to achieve from the crystallization phase in spite of the mutual passage exclusion.


This paper is organized as follows: In Sec. II  we present a model based on an interaction potential for the one dimensional particle system near crystallization. In particular we introduce the simplest version of the Frenkel-Kontoreva model with a flat substrate and a model near crystallization for the paramagnetic interaction between colloidal particles. In Sec. III we provide a revision of the many-particle Langevin equation applied to the one-dimensional system for the harmonic interaction between particles to draw a comparison with the results obtained from the path integral method.  In Sec. IV we present the path integral approach developed for a system in $d-$dimensional space with an effective interaction potential, we show the path integral solution of the $N$-particle Smoluchowski equation. In Sec. V we give the main results for the harmonic approximation obtained from the Langevin description and the path integral approach. Furthermore, starting off the crystal phase we performed subdominant perturbation theory by increasing the temperature and found a clear deviation from the single-file diffusion behavior. In the concluding Sec. VI we give our main remarks and comment on the outlook of this work.

\section{Potential model near crystallization}

The physical system  considered here consists of a set of a brownian interacting particles confined to a straight line. The total interaction potential is assumed to be pairwise additive
\begin{eqnarray}
\Phi\left({\bf x}\right)=\sum_{i<j=1}^{N}V(\left|x_{i}-x_{j}\right|),\label{totalInteraction}
\end{eqnarray}
where $V(r)$ is the pair potential between two particles on the line. Next, we  will consider the one dimensional situation $d=1$, unless stated otherwise and in particular we will consider ${\bf x}=\left(x_{1}, \cdots, x_{N}\right)$ to be an abstract arbitrary position in $\mathbb{R}^{N}$ so that the components of ${\bf x}$ represent the positions of the particles. 

The single file diffusion (SFD) phenomenon in this system is studied close to the crystalline phase  using a harmonization procedure \cite{Lizana2010}.  
Thus it becomes convenient to assume the existence of a mechanical equilibrium point ${\bf x}^{\rm eq}$ such that $\partial_{\ell}\Phi\left({\bf x}^{\rm eq}\right)=0$, with $\ell=1, 2, \cdots, N$,  where the partial derivative is denoted by $\partial_{\ell}:=\frac{\partial}{\partial x_{\ell}}$. Now, it is clear that for any repulsive potential $V(r)$ the particles prefer to be away from each other. In the absence of thermal fluctuations, and if the number of particles remain finite the mechanical equilibrium can be achieved as long as the particles are confined in a finite length $L$, all the while we consider rigid endings; this can be accomplished  physically by anchoring two particles at the end points.  Additionally,  in the thermodynamic limit, $N\to \infty$ and $L\to\infty$, while the particle density $\rho:=N/L$ is kept fixed, this equilibrium can also be guaranteed in the bulk.  As a consequence of constraining the system to a line  the equilibrium positions are found starting off the usual equidistant positions. In particular, for an odd number of particles, these positions would be at 
\begin{eqnarray}
x^{\rm o}_{j}=a(j-1)-\frac{a(N-1)}{2},\label{equilibriumpositions}
\end{eqnarray}
being $a$ the average separation between particles, and also $a=1/\rho$ is approximately the inverse particle density. In the following, we carry out a Taylor expansion around ${\bf x}^{\rm o}$ for the interaction potential to the quadratic order
\begin{eqnarray}
\Phi\left({\bf x}\right)&=&\Phi^{\rm o}-F_{\ell}y^{\ell}+\frac{1}{2}U_{\ell k} ~y^{\ell}y^{k}+\cdots
,\label{TaylorExpansion}
\end{eqnarray}
where the coefficients of the expansion are given by the standard expressions,   $\Phi^{\rm o}:=\Phi\left({\bf x}^{\rm o}\right)$, $F_{j}:=-\partial_{j}\Phi\left({\bf x}^{\rm o}\right)$, $U_{ij}:=\partial_{i}\partial_{j}\Phi\left({\bf x}^{\rm o}\right)$,  and $y^{\ell}:=x_{\ell}-x^{\rm o}_{\ell}$. 
Since, in general $F_{k}\neq 0$, the equilibrium positions in this case are ${x}^{\rm eq}_{i}=x^{\rm o}_{i}+\left(U^{-1}\right)_{ij}F^{j}$.  The coefficients of the Taylor expansion can be obtained using the first and second partial derivatives of the interaction potential, these are given for a pairwise potential function $V(r)$ by 
\begin{eqnarray}
\partial_{k}\Phi\left({\bf x}\right)&=&\sum_{i=1}^{N}V^{(1)}\left(\left|x_{k}-x_{i}\right|\right)\frac{(x_{k}-x_{i})}{\left|x_{k}-x_{i}\right|},\\
\nonumber\\
\partial_{\ell}\partial_{k}\Phi\left({\bf x}\right)&=&\left\{\begin{array}{cc}
\sum_{i=1}^{N}V^{(2)}\left(\left|x_{i}-x_{k}\right|\right), & {\rm for}~ \ell=k, \\
\\
-V^{(2)}\left(\left|x_{\ell}-x_{k}\right|\right), & {\rm for}~ \ell\neq k,
\end{array}\right.
\end{eqnarray} 
where $V^{(1)}(r)$ and $V^{(2)}(r)$ are the first and the second derivative of $V(r)$. 
In the following, we compute the coefficients in (\ref{TaylorExpansion}) shown above for specific interactions potentials.
\vskip1em

\noindent{\bf Example 1}. The simplest potential corresponds to the Frenkel-Kontoreva model with a flat substrate \cite{cha95}. This is given by a quadratic total interaction potential as
\begin{eqnarray}
\Phi\left({\bf x}\right)=\frac{1}{2}\kappa\sum_{i=1}^{N-1}\left(x_{i+1}-x_{i}-l\right)^2,
\label{Harmonic-potencial}
\end{eqnarray}
where the average separation, $a$, between particles in (\ref{equilibriumpositions}) coincides with the  spring's natural length $l$. It is not difficult to show that this potential  can be written as
\begin{eqnarray}
\Phi\left({\bf x}\right)=\frac{\kappa}{2} {\bf x}\cdot\mathbb{V}{\bf x}-{\bf x}\cdot{\bf b}+\frac{1}{2}\kappa l^2\left(N-1\right),
\label{Frenkel-vector}
\end{eqnarray}
where 
${\bf b}=\kappa l\left(-1,0,\cdots, 0,1\right)$ is a constant vector.  Additionally,  this potential corresponds to (\ref{TaylorExpansion}) for $F_{k}=0$,  with $\mathbb{U}=\kappa\mathbb{V}$, where the matrix $\mathbb{V}$ is a tridiagonal matrix whose values are given by the diagonal elements $V_{11
}=V_{NN}=1$, and $V_{kk}=2$ for $k=1, \cdots, N-2$; while the off-diagonal elements $V_{k k+1}=V_{k-1 k}=-1$, and $V_{kk+\ell}=V_{k+\ell k}=0$ for $\ell>0$. 


\vskip1em

\noindent{\bf Example 2}\label{selectionrule}. A  pairwise potential used to compare against experimental results of the SFD phenomenon is the paramagnetic potential \cite{Bechinger2000}, given by
\begin{eqnarray}
V\left(r\right)=\frac{\mu_{0}}{4\pi}\frac{M^2}{r^{3}},\label{ParamagneticPotential}
\end{eqnarray}
where $\mu_{0}$ the vaccum permeability, and  $M$ is  the magnetization of the particles that for paramagnetic colloids satisfies $M=\chi_{\rm eff}B$ being $\chi_{\rm eff}$ an effective magnetic susceptibility and $B$ the transversal applied magnetic field \cite{zahn1997}.

Now, we compute the force acting on the $k-$th particle using $F_{k}:=-\partial_{k}\Phi$ at ${\bf x}^{\rm o}$. By straightforward calculation the force can be expressed as 
\begin{eqnarray}
F_{k}=-\frac{1}{4}\kappa_{P} a\left(\zeta^{(4)}_{k}-\zeta^{(4)}_{N-k+1}\right),
\end{eqnarray}
where $\zeta^{(m)}_{k}:=\zeta\left(m, k\right)$ is the Hurwitz zeta function \cite{Gradshteyn1979} and $\kappa_{P}=3\mu_{0}M^2\rho^{5}/\pi$ is a convenient constant. Clearly, the force on the $k-$th particle is non zero implying that strictly speaking the system is not in mechanical equilibrium at ${\bf x}^{\rm o}$, even in  the absence of thermal fluctuations, reminiscent of the classical Earnshaw's theorem \cite{geim1999magnet}. However, in the large $N$ limit it may be shown that particles in the bulk satisfy $F_{k}\approx 0$. For instance for $N=200$, the particles in positions $x_{35}$ up to $x_{165}$ experiment a force in units of $\kappa_{P} a$ of around $10^{-6}$, meaning that the particles in the bulk are approximately in mechanical equilibrium at ${\bf x}^{\rm o}$. 

Now we proceed to calculate the coefficients of the Taylor expansion (\ref{TaylorExpansion}) at the equidistant positions (\ref{equilibriumpositions}). For the quadratic coefficient one has 
\begin{eqnarray}
U_{\ell k}=\kappa_{P}\left\{\begin{array}{cc}
H_{k-1}^{(5)}+H_{N-k}^{(5)}, & {\rm for}~ \ell=k, \\
\\
-\frac{1}{\left|\ell-k\right|^5}, & {\rm for}~ \ell\neq k,
\end{array}\right. 
\end{eqnarray}
where $H_{n}^{(r)}$ is the harmonic number of order $r$ \cite{Gradshteyn1979}. By numerical inspection $\mathbb{U}=\kappa_{p}\left(\mathbb{V}+\boldsymbol{\epsilon}\right)$, where each element of the matrix $\boldsymbol{\epsilon}$ is less than $3{\rm .}2\times 10^{-2}$.

\section{Langevin approach  revisited: the  harmonic-chain case}

In this section we treat the single file diffusion phenomenon using the many-particle Langevin equation in the overdamped limit.

The mean-square displacement is calculated in order to have a quantity to compare against the result obtained from the  path integral approach.  The Brownian dynamics of the chain can be described  using the overdamped limit of the many-particle Langevin equation
\begin{eqnarray}
 -\gamma \dot{x}_{i}+\sum_{j}F_{ij}+\sqrt{2k_{B}\theta\gamma}\eta_{i}=0,
\label{LangevinEq}
\end{eqnarray}
where $\beta=1/\left(k_{B}\theta\right)$ is the inverse of the thermal energy, being $k_{B}$ Boltzman's constant, and $\theta$ the temperature. Also,  $x_{i}$ represents the $i-$th particle's position, and hydrodynamic interactions amongst the constituents of the chain are neglected.
Here $\eta_{i}$ is a Gaussian stochastic variable with zero mean and variance $1$ for each $i$ such that $1\leq i \leq N$, and, in consistency with the fluctuation-dissipation theorem, these random variables have correlation functions that satisfy
\begin{equation}
    \langle \eta_i (\tau) \eta_j (\tau')  \rangle =\delta_{ij} \delta (\tau- \tau').
\end{equation}
In the above stochastic differential equation (\ref{LangevinEq})  $\gamma$ is a damping friction coefficient,  while $F_{ij}$ denotes the force that the $j$-th particle exerts upon the $i$-th particle. In addition, the free diffusion coefficient is denoted by $D_{0}=k_{B}\theta/\gamma$. In our model at hand all such forces are derived from a potential $\sum_{j}F_{ij}=-\partial_{i}\Phi\left({\bf x}\right)$, where $\Phi({\bf x})$ is the total interaction potential. The average mean-square displacement  $\sigma^{2}_{N}(t)$, is defined as
\begin{eqnarray}
\sigma^{2}_{N}(t)=\frac{1}{N}\sum_{i=1}^{N}
\left(\left<x_{i}^2(t)\right> -\left<x_{i}(t)\right>^2 \right),\label{MSD}
\end{eqnarray}
and can be calculated numerically by a simple Monte-Carlo simulation following Langevin's stochastic description. In the large $N$ limit $\sigma^{2}_{N}\left(t\right)$ is the same as the mean-square displacement of a randomly chosen (tracer) particle in the chain since there is not a preferred particle in the bulk. Using this reasoning, we simply called $\sigma^{2}_{N}\left(t\right)$ the mean-square displacement of the tracer particle. Now we focus on the quadratic expansion of the total interaction potential $\Phi\left({\bf x}\right)=\Phi^{\rm eq}+\frac{1}{2}U_{ij}({\bf x}-{\bf x}^{\rm eq})^{i}({\bf x}-{\bf x}^{\rm eq})^{j}$ in (\ref{TaylorExpansion}).    For this type of potential 
we have $-\nabla\Phi=- \mathbb{U}{\bf x}+{\bf b}$, where ${\bf b}:=\mathbb{U}{\bf x}_{\rm eq}$, and $U_{k \ell}:=\left(\mathbb{U}\right)_{k \ell}$. 
Using these elements  equation (\ref{LangevinEq}) is rendered into a linear Bernoulli equation whose solution in closed form is \cite{Dhont}
\begin{eqnarray}
{\bf x}\left(t\right)=\left<{\bf x}\left(t\right)\right>+\sqrt{2D_{0}}\int_{0}^{t}d\tau e^{-\frac{1}{\gamma}\mathbb{U}\left(t-\tau\right)}\boldsymbol{\eta}\left(\tau\right),
\label{solx}
\end{eqnarray}
while the average value of ${\bf x}(t)$ is
$\left<{\bf x}\left(t\right)\right>=\frac{1}{\gamma}\int_{0}^{t}d\tau e^{-\frac{1}{\gamma}\mathbb{U}(t-\tau)}{\bf b}+e^{-\frac{1}{\gamma}\mathbb{U}t}{\bf x}\left(0\right)$, and $\boldsymbol{\eta}=\left(\eta_{1}, \cdots, \eta_{N}\right)$.
Now using equation (\ref{solx}) it follows directly that
\begin{eqnarray}
\sigma^{2}_{N}(t)=\frac{2D_{0}}{N}\int_{0}^{t}d\tau~{\rm tr}\left(e^{-\frac{2}{\gamma}\mathbb{U}\left(t-\tau\right)}\right)\label{res-langevin}.
\end{eqnarray}
It is noteworthy to mention that this result is valid even if the matrix $\mathbb{U}$ has a zero eigenvalue.



\section{Path integral formulation of Brownian dynamics}\label{PIformulation}


In this section, the path integral approach is introduced in order to study Brownian dynamics. The starting point is the relationship between the Fokker-Plank equation and its path integral formulation developed, for instance, in \cite{Zinn-JustinBook}. This gives  a Feynman path integral representation of the Brownian Dynamics (BD)  of a fluid of strongly interacting particles without hidrodynamic interactions. The following approach is developed in $d$ space dimensions, while below it is applied to one-dimensional systems.  We shall assume that the BD is governed by Smoluchowski equation (see for instance Ref. \cite{Dhont})
 \begin{eqnarray}
\frac{\partial P}{\partial t}=D_{0}\sum_{j=1}^{N}\nabla_{\bf r_{j}}\cdot\left[\beta\left(\nabla_{\bf r_{j}}\Phi\right)P+\nabla_{\bf r_{j}}P\right],\label{Smoluchowski}
\end{eqnarray}
where $P=P\left({\bf r}_{1}, \cdots, {\bf r}_{N},t\right)$  is the probability density function (PDF) of finding the $N$ particles at the positions $({\bf r}_{1}, \cdots, {\bf r}_{N})$ at time $t$, given that at $t_{0}=0$ they were located at $({\bf r}^{\prime}_{1}, \cdots, {\bf r}^{\prime}_{N})$. The above equation (\ref{Smoluchowski}) can be obtained from the Langevin stochastic equation (\ref{LangevinEq}) using standard procedures \cite{GardinerBook}. Although the interaction potential is assumed to be pairwise additive 
$\Phi\left({\bf r}_{1}, \cdots. {\bf r}_{N}\right)=\sum_{i<j=1}^{N}V({\bf r}_{ij})$, wherein ${\bf r}_{ij}={\bf r}_{i}-{\bf r}_{j}$, this condition is not necessary in what follows henceforth. 

It is clearly possible to identify ${\bf x}\in\mathbb{R}^{dN}$ so that ${\bf x}=\left({\bf r}_{1},\cdots,{\bf r}_{N} \right)$ is an abstract general position, and the initial position is denoted by ${\bf x}_{0}=({\bf r}^{\prime}_{1}, \cdots, {\bf r}^{\prime}_{N})$.  Using this identification it is possible to define super-gradient and super-laplacian operators as
\begin{eqnarray}
\boldsymbol{\nabla}&=&(\nabla_{\bf r_{1}}, \cdots, \nabla_{\bf r_{N}}),\nonumber\\
\boldsymbol{\nabla}^{2}&=& \boldsymbol{\nabla} \cdot \boldsymbol{\nabla} =\sum_{j=1}^{N}\nabla^{2}_{\bf r_{j}}.
\end{eqnarray}
In this manner Smoluchowski equation for $N$ interacting particles is identified with Smoluchowski equation for one particle in an Euclidean space of higher dimension $dN$, subjected to an external field $\Phi$, 
\begin{eqnarray}
\frac{\partial P}{\partial t}=D_{0}\boldsymbol{\nabla}^2 P+D_{0}\beta\boldsymbol{\nabla} \cdot\left[\left(\boldsymbol{\nabla}\Phi\right)P\right],
\label{eq}
\end{eqnarray}
where $P\equiv P\left({\bf x}, {\bf x}^{\prime}, t\right)$ is the probability density function to find a particle at the position ${\bf x}$ at time $t$, when at time $t=0$ it was at the position ${\bf x}_{0}$. Additionally,  the initial condition 
$\lim_{t\to 0}P({\bf x}, {\bf x}_{0}, t)=\delta\left({\bf x}-{\bf x}_{0}\right)$,
and the normalization condition $\int_{\mathbb{R}^{dN}} d{\bf x} ~P({\bf x}, {\bf x}_{0}, t)=1$ are  imposed. 

The equation (\ref{eq}) is  a particular example of a Fokker-Planck equation, therefore we may transform it into a Schr\"odinger-like equation, and then look at a path integral corresponding formulation following the same lines implemented in \cite{Zinn-JustinBook, Chaichian, Olindo2009}. Using this procedure, one can prove that the probability density function satisfies 
\begin{eqnarray}
P({\bf x}, {\bf x}_{0}, t)=e^{-\frac{1}{2} \beta\left(\Phi\left({\bf x}\right)-\Phi\left({\bf x}_{0}\right)\right)}\mathbb{P}({\bf x}, {\bf x}_{0}, t), \label{PyP}
\end{eqnarray}
where $\mathbb{P}({\bf x}, {\bf x}_{0}, t)$ satisfies the Schr\"odinger-type equation
$-\frac{\partial \mathbb{P}}{\partial t}=-D_{0}\boldsymbol{\nabla}^2\mathbb{P}+V_{{\rm eff}}\left({\bf x}\right)\mathbb{P}$,
where the resulting ``effective potential'' $V_{\rm eff}\left({\bf x}\right)$, is given by
\begin{eqnarray}
V_{\rm eff}\left({\bf x}\right)=\frac{1}{4}D_{0}\beta^2\left|\boldsymbol{\nabla}\Phi\left({\bf x}\right)\right|^2-\frac{D_{0}\beta}{2}\boldsymbol{\nabla}^{2}\Phi\left({\bf x}\right),\label{effectivepotencial}
\end{eqnarray}
and has the same structure as the one given in \cite{Zinn-JustinBook, Olindo2009}. Using now this Schr\"odinger representation it is possible to write a formal solution to it in terms of a Feynman path integral as follows
 $\mathbb{P}\left({\bf x}, {\bf x}_{0}, t\right)=\int_{{\bf x}_{0}}^{{\bf x}}\mathcal{D}{\bf x}\left(\tau\right)
 e^{-\int_{0}^{t}d\tau\mathcal{L}\left[\dot{\bf x}\left(\tau\right), {\bf x}\left(\tau\right)\right]}$
 where the Lagrangian-like term is given by
 $\mathcal{L}\left[\dot{\bf x}\left(\tau\right), {\bf x}\left(\tau\right)\right]=\frac{1}{4D_{0}}\dot{\bf x}^2\left(\tau\right)+V_{\rm eff}\left({\bf x}\left(\tau\right)\right)$,
and $\mathcal{D}{\bf x}$ is an appropriate functional measure. Furthermore,  one has the following identity $\Phi\left({\bf x}\right)-\Phi\left({\bf x}_{0}\right)=\int_{{\bf x}_{0}}^{{\bf x}} d\tau~\dot{\bf x}\left(\tau\right)\cdot\nabla\Phi\left({\bf x}\left(\tau\right)\right)$. Hence,  using (\ref{PyP})   a formal solution in the path integral representation for the Smoluchowski probability density function $P\left({\bf x}, {\bf x}_{0}, t\right)$ can be obtained as
\begin{eqnarray}
P\left({\bf x}, {\bf x}_{0}, t\right)=\int_{{\bf x}_{0}}^{{\bf x}}\mathcal{D}{\bf x}\left(\tau\right)
 e^{-\int_{0}^{t}d\tau L\left[\dot{\bf x}\left(\tau\right), {\bf x}\left(\tau\right)\right]},\label{Smoluchowskiopath}
\end{eqnarray}
where the Lagrangian-like function $L$ is given by 
\begin{eqnarray}
L\left[\dot{\bf x}(\tau), {\bf x}(\tau)\right]=\frac{\gamma\beta}{4}\left(\dot{\bf x}\left(\tau\right)+\frac{1}{\gamma}\boldsymbol{\nabla}\Phi\left({\bf x}\left(\tau\right)\right)\right)^{2}-\frac{1}{2\gamma}\boldsymbol{\nabla}^2\Phi\left({\bf x}\left(\tau\right)\right),\label{Lagrange}
\end{eqnarray}
here the relation $D_{0}=1/(\gamma\beta)$ was used, recalling that  $\gamma$ is the damping friction coefficient. 

The path integral representation (\ref{Smoluchowskiopath}) shall be the starting point to our approach to the SFD. In particular, as expected, exact analytical results are obtained for the harmonic potential (\ref{Harmonic-potencial}) using this method. In addition, approximated results near the crystallization phase can be reached using other more realistic potentials. This latter point will be illustrated using a paramagnetic potential.




For most potentials the above path integral cannot be solved exactly, and approximation methods are most useful, for instance the method of steepest descent. Usually this procedure works best in the weak potential regime \cite{Chaichian}, here  however it is appropriate for the low temperature regime so as to maintain strong interactions and capture the SFD behaviour. 
In the Lagrangian (\ref{Lagrange}) the second term $-\boldsymbol{\nabla}^{2}\Phi/(2\gamma)$ does not depend on temperature,  therefore it can be treated as a perturbation. The basic idea behind this technique consists of expanding the Lagrangian (\ref{Lagrange}) around a classical solution of the corresponding Euler-Lagrange equations $\frac{d}{d\tau}\left(\frac{\partial L}{\partial \dot{\bf x}}\right)=\frac{\partial L}{\partial {\bf x}}$, namely 
\begin{eqnarray}
\frac{\beta\gamma}{2}\ddot{\bf x}=\boldsymbol{\nabla}V_{\rm eff}\left({\bf x}\right).\label{E-L}
\end{eqnarray}
Let us call ${\bf  x}_{\rm cl}\left(\tau\right)$ the classical solution of the Euler-Lagrange equations (\ref{E-L}) satisfying the boundary conditions ${\bf x}_{0}={\bf x}_{\rm cl}(0)$ and ${\bf x}={\bf x}_{\rm cl}(t)$, where ${\bf x}_{0}$ and ${\bf x}$ are the initial and final positions in the path integral configuration (\ref{Smoluchowskiopath}). 

Now, we write the trajectory as ${\bf x}={\bf x}_{\rm cl}+{\bf q}$, where ${\bf q}$ is interpreted as a fluctuation around the classical solution.
The boundary conditions chosen imply for the fluctuation ${\bf q}\left(0\right)={\bf q}(t)=0$. Taking this into account the PDF can be rewritten as 
\begin{eqnarray}
P\left({\bf x}, {\bf x}_{0}, t\right)&=&e^{-\int_{0}^{t}d\tau L\left[\dot{\bf x}_{\rm cl}\left(\tau\right), {\bf x}_{\rm cl}\left(\tau\right)\right]}
\mathcal{P}\left({\bf x}, {\bf x}_{0}, t\right),\label{genericPath}
\end{eqnarray}
where the function $\mathcal{P}\left({\bf x}, {\bf x}_{0}, t\right)$ is given by
\begin{eqnarray}
\mathcal{P}\left({\bf x}, {\bf x}_{0}, t\right)=\oint \mathcal{D}{\bf q}\left(\tau\right)
e^{-\int_{0}^{t}d\tau\left(L\left[\dot{\bf x}_{\rm cl}\left(\tau\right)+\dot{\bf q}\left(\tau\right), {\bf x}_{\rm cl}\left(\tau\right)+{\bf q}\left(\tau\right)\right]-L\left[\dot{\bf x}_{\rm cl}\left(\tau\right), {\bf x}_{\rm cl}\left(\tau\right)\right]\right)}.\label{intoverq}
\end{eqnarray}
The next step consists in expading $L\left[\dot{\bf x}_{\rm cl}+\dot{\bf q}, {\bf x}_{\rm cl}+{\bf q}\right]$ in powers of ${\bf q}\left(\tau\right)$, and evaluate (\ref{intoverq}) approximately. 


 \section{Single file diffusion from the wordline approach }

 In this section we present applications of the path integral approach to SFD in concrete models. 

\subsection{Probability density function $P\left({\bf x}, {\bf x}_{0}, t\right)$ for the quadratic model.} \label{probsect}


The aim of this section is to determine the probability density function $P\left({\bf x}, {\bf x}_{0}, t\right)$ using the path integral formulation of brownian dynamics, (\ref{Smoluchowskiopath}), presented in the last section. Let us consider a system consisting of $N$ bodies interacting  through the potential (\ref{totalInteraction}) embedded in a fluid whose molecules have geometric dimensions that are much smaller than those of the bodies. This system corresponds to a finite brownian chain. In the thermodynamic limit, when $N \to \infty$, we shall simply refer to the system as a brownian  chain. 

Clearly, it is not possible to treat the general situation for any pairwise potential $V(r)$, however, let us assume the existence of the mechanical equilibrium ${\bf x}^{\rm eq}$ considered in the last subsection. This equilibrium can be achieved classically in the low temperature regime, and hence as far as the damping friction parameter $\gamma$ is kept fixed, the steepest descent method can be carried out in  (\ref{Smoluchowskiopath}) as long as $\beta$ is large. A manner to proceed is to start off from 
 the quadratic approximation (\ref{TaylorExpansion}),  where we have $-\boldsymbol{\nabla}\Phi\left({\bf x}\right)=-\mathbb{U}\left({\bf x}-{\bf x}^{\rm eq}\right)$ and $-\boldsymbol{\nabla}^2\Phi=-{\rm tr}\left(\mathbb{U}\right)$. We find that the effective potential $V_{\rm eff}\left({\bf x}\right)$ (\ref{effectivepotencial}) is in fact
\begin{eqnarray}
V_{\rm eff}\left({\bf x}\right)=\frac{1}{4}D_{0}\beta^2\left[\mathbb{U}\left({\bf x}-{\bf x}^{\rm eq}\right)\right]^2-\frac{1}{2}D_{0}\beta{\rm tr}\left(\mathbb{U}\right).
\end{eqnarray}
It is now convenient to introduce the change of variables ${\bf y}={\bf x}-{\bf x}_{\rm eq}$. In this manner the path integral appearing in $P$, (\ref{Smoluchowskiopath}), may be rewritten as 
\begin{eqnarray}
P({\bf y}, {\bf y}_{0}, t)=\int_{{\bf y}_{0}}^{{\bf y}}\mathcal{D}{\bf y}\left(\tau\right)
e^{-\int_{0}^{t}d\tau L\left[\dot{\bf y}\left(\tau\right), {\bf y}\left(\tau\right)\right]},
\label{path}
\end{eqnarray}
where 
the Lagrangian function is 
\begin{eqnarray}
L\left[\dot{\bf y}(\tau), {\bf y}(\tau)\right]=\frac{\gamma\beta}{4}\left(\dot{\bf y}\left(\tau\right)+\frac{1}{\gamma}\mathbb{U}{\bf y}\left(\tau\right)\right)^{2}-\frac{1}{2\gamma}{\rm tr}\left(\mathbb{U}\right).
\label{LagrangianG0}
\end{eqnarray}
Now, since  the matrix $\mathbb{U}$ is diagonalizable 
there exists an  $N\times N$ orthogonal matrix, $\mathcal{O}$, such that $\mathcal{O}\mathbb{U}\mathcal{O}^{\dagger}={\rm diag}(\lambda_{1}, \cdots, \lambda_{N})$ is a diagonal matrix, where $\left\{\lambda_{i}\right\}_{i=1, \cdots, N}$ are its eigenvalues. After the coordinate transformation ${\bf y}=\mathcal{O}{\bf z}$, the path integral (\ref{path}) becomes
\begin{eqnarray}
P({\bf z}, {\bf z}_{0}, t)=\prod_{\ell=1}^{N}\int_{{\bf z}_{0}}^{{\bf z}}\mathcal{D}{z}_{\ell}\left(\tau\right)
e^{-\int_{0}^{t}d\tau L_{\ell}\left(\dot{z}_{\ell}\left(\tau\right), z_{\ell}\left(\tau\right)\right)},
\label{path2}
\end{eqnarray}
where the Lagrangian $L_{\ell}$ for the $\ell$-th mode is given by 
\begin{eqnarray}
L_{\ell}\left(\dot{z}_{\ell}, z_{\ell}\right)=\frac{1}{4D_{0}}\left(\dot{z}_{\ell}\left(\tau\right)+\omega_{\ell}{z}_{\ell}\left(\tau\right)\right)^{2}-\frac{1}{2}D_{0}\beta{\rm tr}\left(\mathbb{U}\right), \label{Lagrangian0}
\end{eqnarray}
where $\omega_{\ell}=D_{0}\beta \lambda_{\ell}$ are the normal frequencies.   
In addition, for completeness, the Euler-Lagrange equation, $\ddot{z}_{\ell}=\omega_{\ell}z_{\ell}$, for the Lagrangian in (\ref{path2}) 
has the following classical solution
\begin{eqnarray}
z^{\ell}_{\rm cl}\left(\tau\right)=\frac{z_{\ell}\sinh\left(\omega_{\ell}\tau\right)+z^{\ell}_{0}\sinh\left(\omega_{\ell}(t-\tau)\right)}{\sinh\left(\omega_{\ell}t\right)},\label{classicalsol}
\end{eqnarray}
where $z^{\ell}_{0}:=z^{\ell}(0)$ and $z^{\ell}:=z^{\ell}(t)$ are the initial and final positions of the $\ell$ mode. Notice that (\ref{Lagrangian0}) has the same Euler-Lagrange equations as the harmonic oscillator with imaginary time.  
 Additionally, each one of the path integrals appearing in $P({\bf z}, {\bf z}_{0}, t)$ is closely related to the path integral corresponding to a quantum harmonic oscillator (QHO) after the identification of parameters: $m=\frac{1}{2D_{0}}$, and  $\omega/\hbar=D_{0}\beta\lambda_{k}$ where $m$ is the mass and $\omega$ the frequency of the QHO, respectively, and the imaginary time correspondence $t=-\frac{i}{\hbar}t_{QHO}$, where $t_{QHO}$ is the time parameter in the QHO. The path integral of the QHO has been solved exactly (see e.g. \cite{Zinn-JustinBook, Chaichian}), it is then not surprising to find an exact result for (\ref{path2}). Now, the path integral (\ref{genericPath}), in this case, is obtained by substituting the above classical solution in the Lagrangian (\ref{Lagrangian0}), and then using  (\ref{intoverq}), this latter quantity turns out to be independent on $z_{0\ell}$ and $z_{\ell}$, indeed, one has $\mathcal{P}\left(z_{\ell}, z_{0\ell}, t\right):=\mathbb{Z}$, where
 \begin{eqnarray}
\mathbb{Z}=\oint \mathcal{D}q_{\ell}\left(\tau\right)e^{-\int_{0}^{t}d\tau \frac{1}{4D_{0}}\left(\dot{q}_{\ell}\left(\tau\right)+\omega_{\ell}q_{\ell}\left(\tau\right)\right)^{2}},\label{PartitionF}
 \end{eqnarray}
 depends only on time $t$. Since $P\left({\bf x}, {\bf x}_{0}, t\right)$ is requested to be normalized the value of $\mathcal{P}\left(z_{\ell}, z_{0\ell}, t\right)$ satisfies a normalization condition. 
 Considering this condition, 
 the Smoluchowski probability density (\ref{Smoluchowskiopath}) is found straightforwardly to be
\begin{eqnarray}
P({\bf z}, {\bf z}_{0}, t)&=&\prod_{\ell=1}^{N}\frac{1}{\sqrt{2\pi \sigma^{2}_{\ell}}}
e^{-\frac{\left(z_{\ell}-z^{\prime}_{0\ell}\right)^2}{2\sigma^{2}_{\ell}}},
\label{path3}
\end{eqnarray}
%
where $\sigma^{2}_{\ell}$ and $z^{\prime}_{0\ell}$ are  given by
\begin{eqnarray}
\sigma^{2}_{\ell}&=&\frac{2D_{0}\left(\coth\left(\omega_{\ell}t\right)+1\right)^{-1}}{\omega_{\ell}},\label{indVar}\\
z^{\prime}_{0\ell}&=&e^{-\omega_{\ell} t}z_{0\ell}.
%
\end{eqnarray}
Notice that the matrix $\mathbb{U}$ may have zero modes, in that case we can split the degrees of freedom with zero modes from those with nonzero modes in (\ref{path2}). An equivalent regulating procedure is to add  the diagonal matrix $\delta {\bf 1}$ to $\mathbb{U}$, and after mean-values are evaluated, take the limit $\delta\to 0^+$; this limit can also be taken directly in the probability density function to evaluate it. 
 
\subsection{Mean-square displacement.}

In this section we will use the PDF (\ref{path3}) obtained previously to calculate the mean square displacement (MSD) (\ref{MSD}). First notice that $\sigma^{2}_{N}(t)$, in the quadratic approximation,  is invariant under the transformation ${\bf x}\to{\bf z}=\mathcal{O}\left({\bf x}-{\bf x}_{\rm eq}\right)$, which entails $\sigma^{2}_{N}(t)=\frac{1}{N}\left<\left({\bf z}-\left<{\bf z}\right>\right)^2\right>$; this quantity corresponds exactly to the variance of $P({\bf z}, {\bf z}_{0}, t)$ (\ref{path3}), and from (\ref{indVar}) it follows that
\begin{eqnarray}
\sigma^{2}_{N}(t)=\frac{1}{N}\sum_{\ell=1}^{N}\sigma^{2}_{\ell}=\frac{1}{N}\sum_{\ell=1}^{N}\frac{2D_{0}\left(\coth\left(\omega_{\ell}t\right)+1\right)^{-1}}{\omega_{\ell}}.\label{MSD-HArm}
\end{eqnarray}
Using the definition of $\coth(x)$ it is not difficult to see that this expression (\ref{MSD-HArm}) is indeed the same as the one found using the Langevin equation (\ref{res-langevin}). Therefore (\ref{MSD-HArm}) demonstrates the effectiveness of the proposed  Feynman path integral approach to solve Smoluchowski equation within the physical assumptions in our working example.


\subsection{Improved expression for the mean-square displacement.}

In this section we present an improvement for the mean-square displacement. The essential idea is to expand up to quartic order the total interaction potential $\Phi({\bf x})$ around an equilibirum ${\bf x}_{\rm eq}$, keeping a quadratic order approximation for the Lagrangian (\ref{Lagrange}). The Taylor expansion of $\Phi({\bf x})$ up to quartic order in ${\bf y}:={\bf x}-{\bf x}_{\rm eq}$ is
\begin{eqnarray}
\Phi\left({\bf x}\right)\simeq\Phi_{\rm eq}+\frac{1}{2}U_{\ell k} ~y^{\ell}y^{k}+\frac{1}{3!}U^{(3)}_{\ell k m} ~y^{\ell}y^{k}y^{m}+\frac{1}{4!}U^{(4)}_{\ell k m n} ~y^{\ell}y^{k}y^{m}y^{n},\label{TaylorExpansion22}
\end{eqnarray}
where coefficients $U^{(3)}_{\ell k m}$, and $U^{(4)}_{\ell k m n}$ are obtained in \ref{apppre22}. Keeping quadratic terms in (\ref{Lagrange}) is  tantamount to the following replacements for $\boldsymbol{\nabla} \Phi$ and $\boldsymbol{\nabla}^{2}\Phi$
\begin{eqnarray}
\boldsymbol{\nabla}^{2}\Phi &\simeq&{\rm tr}\mathbb{U}+U^{(3)}_{\ell\ell k}y^{k}+\frac{1}{2}U^{(4)}_{\ell\ell ij}y^{i}y^{j},\nonumber\\
\left(\boldsymbol{\nabla}\Phi\right)_{i}&\simeq& U_{ij}y^{j}.
\end{eqnarray}
The Lagrangian (\ref{Lagrange}) is now given approximately by 
\begin{eqnarray}
L\left[\dot{\bf x}(\tau), {\bf x}(\tau)\right]=\frac{\gamma\beta}{4}\left(\dot{\bf y}+\frac{1}{\gamma}\mathbb{U}{\bf y}\right)^{2}-\frac{1}{2\gamma}{\rm tr}\mathbb{U}
-\frac{1}{2\gamma}V_{I}({\bf y}).
\label{Lagrange2}
\end{eqnarray}
where $V_{I}({\bf y})=U^{(3)}_{\ell\ell k}y^{k}+\frac{1}{2}U^{(4)}_{\ell\ell ij}y^{i}y^{j}$. The first two terms in the above Lagrangian correspond naturally to the approximation (\ref{LagrangianG0}) treated in the preceeding subsection. Most importantly, in the low-temperature regime the term $V_{I}({\bf y})$ can be handled as a perturbative term.  

After the change of coordinates ${\bf y}=\mathcal{O}{\bf z}$ we obtain
\begin{eqnarray}
P({\bf z}, {\bf z}_{0}, t)=\int_{{\bf z}_{0}}^{{\bf z}}\prod_{\ell=1}^{N}\mathcal{D}{z}_{\ell}\left(\tau\right)
e^{-\int_{0}^{t}d\tau \left[\sum_{\ell=1}^{N}L_{\ell}\left(\dot{z}_{\ell}\left(\tau\right), z_{\ell}\left(\tau\right)\right)-\frac{1}{2\gamma}\overline{V}_{I}(z)\right]},
\label{path4}
\end{eqnarray}
where $\overline{V}_{I}({\bf z})=V^{(1)}_{k}z^{k}+\frac{1}{2}V^{(2)}_{ij}z^{i}z^{j}$ is the transformed perturbative potential in the new coordinates, being $V^{(1)}_{k}=\mathcal{O}_{j k}U_{\ell\ell j}$, and $V^{(2)}_{kk^{\prime}}=\mathcal{O}_{ik}\mathcal{O}_{ik^{\prime}}U^{(4)}_{\ell\ell ij}$. Next we expand the exponential containing the perturbation term, we use the steepest descent approximation, and then exponentiate it back again. It is not difficult to prove that 
\begin{eqnarray}
P({\bf z}, {\bf z}_{0}, t)\simeq \frac{1}{\mathcal{Z}}e^{-\int_{0}^{t}d\tau \sum_{\ell=1}^{N}\frac{1}{4D_{0}}\left(\dot{z}^{\ell}_{\rm cl}\left(\tau\right)+\omega_{\ell}{z}^{\ell}_{\rm cl}\left(\tau\right)\right)^{2}+\frac{1}{2\gamma}\int_{0}^{t}d\tau \left<\overline{V}_{I}\left({\bf z}_{\rm cl}(\tau)+{\bf q}(\tau)\right)\right>},
\label{path4}
\end{eqnarray}
where $\mathcal{Z}$ is the normalization factor, and $z^{\ell}_{\rm cl}(\tau)$ is the classical solution (\ref{classicalsol}). Additionally, the expectation value $\left<\cdots\right>$ is found as usual
\begin{eqnarray}
\left<\cdots\right>=\frac{1}{\mathbb{Z}}\oint \mathcal{D}{\bf q}\left(\tau\right)\cdots
e^{-\int_{0}^{t}d\tau\frac{1}{4D_{0}}\left(\dot{\bf q}\left(\tau\right)+\frac{1}{\gamma}\mathbb{U}{\bf q}\left(\tau\right)\right)^{2}},
\label{defaverage}
\end{eqnarray}
while the partition function is given by (\ref{PartitionF}). 

Next, we will obtain an expression for the joint-probability function $P({\bf z}, {\bf z}_{0}, t)$ computing the expectation value $\left<\overline{V}_{I}\left({\bf z}_{\rm cl}(\tau)+{\bf q}(\tau)\right)\right>$ (see \ref{applast}). After performing the $\tau$ integrations in (\ref{path4}) it can be shown that
\begin{eqnarray}
P\left({\bf z}, {\bf z}_{0}, t\right)=\frac{1}{\mathcal{Z}}
e^{-\frac{1}{2}\sum_{\ell, k}z_{\ell}A_{\ell k}z_{k}+\sum_{\ell}J_{\ell}z_{\ell}},
\label{PDFnonharmonic}
\end{eqnarray}
where the partition function $\mathcal{Z}$, associated to the joint density probability function is given by 
\begin{eqnarray}
\mathcal{Z}=\int \left(\prod_{\ell=1}^{N}dz^{\ell}\right)
e^{-\frac{1}{2}\sum_{\ell, k}z_{\ell}A_{\ell k}z_{k}+\sum_{\ell}J_{\ell}z_{\ell}},
\label{newPF}
\end{eqnarray}
with the coefficients
\begin{eqnarray}
A_{\ell k}&=&\frac{\delta_{\ell k}}{\sigma^{2}_{\ell}}-\frac{1}{4\gamma}V^{(2)}_{\ell k}\mu_{\ell k},\\
J_{k}&=&\frac{z^{\prime}_{0k}}{\sigma^{2}_{k}}+\frac{1}{4\gamma}V^{(1)}_{k}\nu_{k}-\frac{1}{2\gamma}\sum_{\ell}V^{(2)}_{k\ell}z_{0}^{\ell}\eta_{\ell k},
\end{eqnarray}
where we defined $\mu_{\ell k}=(\omega_{\ell}\coth\left(\omega_{\ell}t\right)-\omega_{k}\coth\left(\omega_{k}t\right))/(\omega^{2}_{\ell}-\omega^{2}_{k})$, $\nu_{k}=\tanh\left(\frac{\omega_{k}t}{2}\right)/\omega_{k}$ and $\eta_{\ell k}=(\omega_{\ell}{\rm csch}\left(\omega_{\ell}t\right)-\omega_{k}{\rm csch}\left(\omega_{k}t\right))/(\omega^{2}_{\ell}-\omega^{2}_{k})$. 

Note that $P({\bf z},{\bf z}_{0}, t)$ in (\ref{PDFnonharmonic}) reduces to (\ref{path3}) when ${V}^{(1)}$ and ${V}^{(2)}$ are neglected as it should.

It is well known that the partition function $\mathcal{Z}$ can be solved in closed form \cite{Zinn-JustinBook}
\begin{eqnarray}
\mathcal{Z}=\left(\sqrt{2\pi}\right)^{N}\left(\det\left(\mathbb{A}\right)\right)^{-1/2}
e^{\frac{1}{2}\sum_{ij}J_{i}\left(A^{-1}\right)_{ij}J_{j}}.
\end{eqnarray}
Using this expression for $\mathcal{Z}$ one can generate all the correlation functions using the standard procedure, that is, for the correlation functions 
$\left<z_{i_{1}}\cdots z_{i_{n}}\right>=\frac{\partial}{\partial J_{i_{1}}}\cdots\frac{\partial}{\partial J_{i_{n}}}\mathcal{Z}$, 
while the connected correlation functions can be obtained similarly using the  formula
$\left<z_{i_{1}}\cdots z_{i_{n}}\right>_{c}=\frac{\partial}{\partial J_{i_{1}}}\cdots\frac{\partial}{\partial J_{i_{n}}}\log\mathcal{Z}$. 
For instance, the mean-square displacement can be calculated using
\begin{eqnarray}
\sigma_{N}^{2}\left(t\right)=\frac{1}{N}\sum_{i=1}^{N}\left<z^{2}_{i}\right>_{c}.\label{MSDc}
\end{eqnarray}



For the MSD it is enough consider the approximation $\left(A^{-1}\right)_{\ell k}\simeq \sigma^{2}_{\ell}\delta_{\ell k}+\frac{1}{4\gamma}\sigma^{2}_{\ell}\left(V^{(2)}_{\ell k}\mu_{\ell k}\right)\sigma^{2}_{k}$. Taking into account this approximation, the MSD can be cast into the form
\begin{eqnarray}
\sigma_{N, {\rm imp}}^{2}\left(t\right)&=&\frac{1}{N}\sum_{\ell}\sigma^{2}_{\ell}\left\{1+\frac{1}{4\gamma}\sigma^{2}_{\ell}\mu^{2}_{\ell}V^{(2)}_{\ell\ell}+\cdots\right\},\label{MSD-corrections}
\end{eqnarray}
where $\mu^{2}_{\ell}=\left(\coth\left(\omega_{\ell}t\right)-\left(\omega_{\ell}t\right){\rm csch}^{2}\left(\omega_{\ell}t\right)\right)/\omega_{\ell}$
The first term of this expression corresponds to the MSD of the quadratic approximation  (\ref{MSD-HArm}). 

For the harmonic approximation as well as the present approximation, there are two situations to be considered, the case when $N$ is finite, and the thermodynamic limit when $N\to\infty$.

\section{Results and discussion}

\subsection{Calculation of MSD for quadratic potential}\label{Analytics}
Here, we will carry out the calculation of the mean-square displacement using the expression (\ref{MSD-HArm}) valid for a potential given by $\Phi\left({\bf x}\right)=\Phi^{\rm eq}+\frac{1}{2}U_{ij}({\bf x}-{\bf x}^{\rm eq})^{i}({\bf x}-{\bf x}^{\rm eq})^{j}$ (\ref{TaylorExpansion}). This potential is the same as the harmonic potential when $\mathbb{U}=\kappa \mathbb{V}$, and near the crystalline phase this potential approximates the paramagnetic potential for nearest-neighbor interactions when $\mathbb{U}\simeq\kappa_{P}\mathbb{V}$. Therefore the model becomes a valid treatment for both harmonic interacting particles and paramagnetic particles near crystallization.  

This harmonic interaction potential introduces naturally a temporal scale associated with overdamping given by $t_{0}=\gamma/\kappa$, or $\gamma/\kappa_{P}$ for the paramagnetic potential. 

Remarkably an analytical result using (\ref{MSD-HArm}) can be found by calculating the eigenvalues $\lambda_{\ell}$ of the tridiagonal matrix $\mathbb{V}$,  
where $\lambda_{\ell}$ are the
 eigenvalues of $\mathbb{V}$, given by 
$\lambda_{\ell}=4\sin^2\left(\frac{\left(\ell-1\right)\pi}{2N}\right)$
with $\ell=1,\cdots, N$, where $\ell=1$ corresponds to the zero mode (see \ref{B}).

Since there is a zero eigenvalue it is convenient to split the sum into $\ell=1$ and $\ell\geq 2$, and in considering the zero mode it is convenient to regulate such mode, as described in subsection \ref{probsect}, indeed $\lambda_{1}(\delta)= \delta$ and then the limit $\delta\to 0$ is taken afterwards to find\footnote{We may also express formally $\sigma^{2}_{N}(t)=\frac{D_{0}t_{0}}{N}\sum_{\ell=1}^{N}\left[\frac{1-e^{-2\lambda_{\ell}t/t_{0}}}{\lambda_{\ell}}\right]$ interpreting the singular term using L'Hospital's rule and arrive at the same result.}
\begin{eqnarray}
\sigma^{2}_{N}(t)=\frac{2D_{0}t}{N}+\frac{D_{0}t_{0}}{N}\sum_{\ell=2}^{N}\left[\frac{1-e^{-2\lambda_{\ell}t/t_{0}}}{\lambda_{\ell}}\right].\label{res}
\end{eqnarray}
The first term in this expression represents the usual diffusive behavior with a reduced diffusion coefficient $D_{0}/N$ and is to be interpreted as the free diffusion of the whole cluster of particles \cite{Delfau2011}.  

The short- and long-time regimes are defined by $t\ll t_{0}$ and $t\gg t_{0}$, respectively, the formula (\ref{res}) has the asymptotic corresponding behaviors
\begin{eqnarray}
\sigma^{2}_{N}\left(t\right)=\left\{\begin{array}{cc}
2D_{0}t,     & {\rm for}~~ t\ll t_{0}, \\
\\
\frac{2D_{0}}{N}t,     & {\rm for }~~ t\gg t_{0}. 
\end{array}\right.
\label{asymptotic}
\end{eqnarray}
This result establishes that, as long as $N$ remain finite,  there are two diffusive regimes in the line: the free diffusion regime with diffusion coefficient $D_{0}$ and a cluster diffusion regime with reduced diffusion coefficient $D_{N}:=D_{0}/N$ \cite{Delfau2011} and, on the other hand, an explicit behavior of the MSD in the thermodynamic limit $N\to\infty$. To analyse the thermodynamic limit $N\to\infty$ of MSD, we approximate the sum in (\ref{res}) by an integral $\sigma^{2}\left(t\right):=\lim_{N\to\infty}\sigma^{2}_{N}\left(t\right)$, one has
\begin{eqnarray}
\sigma^{2}(t)= \frac{D_{0}t_{0}}{4}\int_{0}^{1}dp\frac{1-e^{-8\sin^2\left(\frac{\pi p}{2} \right)T }}{\sin^2\left(\frac{\pi p}{2}\right)},\label{integral}
\end{eqnarray}
where $T=t/t_{0}$. We now expand the exponential function, and after identity (\ref{id}) we find the following series representation 
\begin{eqnarray}
\sigma^{2}(t)=\frac{D_{0}t_{0}}{4\sqrt{\pi}}\sum_{k=1}^{\infty}\left(-1\right)^{k+1}\frac{\left(8T\right)^{k}}{k!}\frac{\Gamma\left(k-\frac{1}{2}\right)}{\Gamma\left(k\right)},
\label{series-rep}
\end{eqnarray}
which is most useful in the short-time behavior. This series can also summed in terms of Bessel functions, 
to this end consider the change of variables $\theta = \pi p$,  and the derivative of the integral (\ref{integral}) $
d\sigma^{2}/dT
= 8D_{0}t_{0} e^{-4T} I_0 (4T)
$, where the classical integral representation of the Bessel function, $I_0 (x) = \frac{1}{\pi} \int_0^{\pi} d\theta e^{x \cos \theta}$, was used. Finally, integration over $T$ can be carried out integrating by parts and using the basic identities (\ref{id2}) fixing the integration constant from $\sigma^{2}(0)=0$.  

The mean-square displacement turns out to be
\begin{eqnarray}
\sigma^{2}(t)=2D_{0}t_{0}T~e^{-4T}\left[I_{0}\left(4T\right)+I_{1}\left(4T\right)\right], \label{res2}
\end{eqnarray}
where $I_{k}(x)$ are the modified Bessel functions of the first kind and order $k
$, from which the series representation  (\ref{series-rep}) can also be obtained. From the asymptotic of the Bessel functions we find that the $\sigma^{2}(t)$ behaves in the long- and short-time regimes as 
\begin{eqnarray}
\sigma^{2}(t)\approx 2D_{0}\left\{\begin{array}{cc}
\left(\frac{t_{0}}{2\pi}\right)^{\frac{1}{2}} t^{\frac{1}{2}},&~~~{\rm for} ~~t\gg t_{0}, \\
\\
t, &~~~{\rm for }~~ t\ll t_{0}.
\end{array}\right.
\label{sfd}
\end{eqnarray}
For $t\ll t_{0}$ the systems presents the diffusive behaviour of a free particle, and for $t\gg t_{0}$ it has the typical behavior of the single-file diffusion. 

The approximation obtained has the same asymptotic behavior as that reported in \cite{Lizana2010}. Using both behaviors it is possible to determine a time transition between both regimes by equating the short- and long-time behavior, we estimate such transition time to be $t_{*}=\frac{k_{B}\theta}{2D_{0}\pi\kappa}$.
Now, in the harmonic and nearest-neighbour approximation the system of paramagnetic particles, we find
\begin{eqnarray}
t_{*}\simeq \frac{k_{B}\theta}{6D_{0}\mu_{0}M^2 \rho^5},
\end{eqnarray} 
showing that the transition time is shorter when the particle density (or magnetization) is larger. 

%

\subsubsection{ Numerical analysis of MSD.}\label{numerics}

Consider the dimensionless quantities $\mu_{N}(t) := \sigma^{2}_{N}(t)/D_0 t_0$, $T=t/t_0$ and perform the change of variables $\tau = T( 1-u)$, and $\boldsymbol{\zeta} (u) := \sqrt{T} \boldsymbol{\eta} (T(1-u))$
we can then calculate $\sigma^{2}(t)$ independently of the length scale by the numerical estimate of the average
\begin{equation}
   \mu_{N}(T) = \frac{2T}{N} \int_0^1\!\! du\!\! \int_0^1 du'
    \langle \boldsymbol{\zeta }(u') \cdot e^{-T(u+u')\mathbb{V}}\boldsymbol{ \zeta} (u) \rangle,
\end{equation}
through generation of large samples of random numbers with the required normal Gaussian distribution and correlation functions to approximate the relation $\langle \zeta_i (u) \zeta_j (u')  \rangle =\delta_{ij} \delta (u- u').$
To this end we partition the unit interval into $M$ equal size subintervals so that the integration variable becomes discretized $u\to \alpha/M$ and $\int_0^1 du f(u) \to \frac{1}{M} \sum_{\alpha=1}^M f(\alpha /M)$. In particular we will replace $\zeta_i (u) \to \zeta_i^{\alpha} = \sqrt{M} z_i^{\alpha}$ where each $z_i^\alpha$ is a stochastic random variable normally distributed with zero mean and unit variance. The correlation function now takes the form $\langle \zeta_i^\alpha \zeta_j^\beta \rangle = M \delta_{ij}\delta_{\alpha \beta}$. We shall further introduce collective indices $A = (i,\alpha)$ ranging $1 \leq A \leq MN$

\begin{eqnarray}
  \mu_{N}(T) &\approx& \frac{2T}{MN} \sum_{\alpha,\beta =1}^{M} \sum_{i,j=1}^{N} \langle z_i^\alpha z_j^\beta \rangle [e^{-T(\alpha + \beta)\mathbb{V}/M}]_{ij}\nonumber \\ \label{mu}
& =& \frac{2T}{NM} \sum_{A,B =1}^{MN}Z_{AB} K_{AB},
\end{eqnarray}
where we have introduced the matrix representation $K_{AB}$ of the kernel $[exp(-T(\alpha + \beta)\mathbb{V}/M)]_{ij}$ traced against the matrix representation $Z_{AB} = Z_{BA}$ of the average $\langle z_i^\alpha z_j^\beta \rangle$. 
In figure 1 we display and compare the numerical results against their analytical counterpart for $\mu_{N}(T)$ for different number of particles.
\begin{figure}[h!]
\begin{center}
\includegraphics[scale=1]{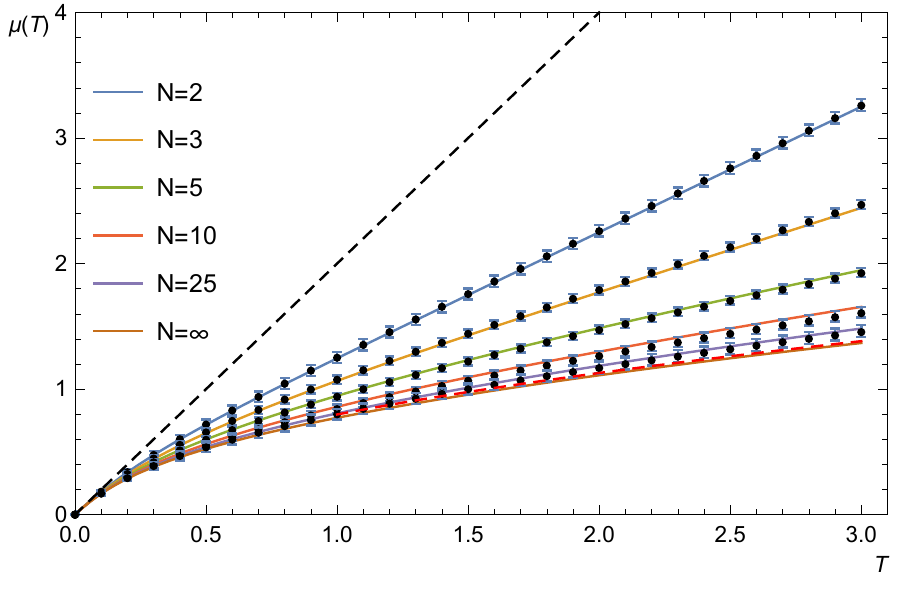}
\caption{{ \small Asymptotes $2T, \sqrt{2T/\pi}$ are shown as dashed lines. The thermodynamic limit $N \to \infty$ is shown in brown. The points are numerical calculations of $\mu(T)$ for different number of particles $N$ while solid lines are their analytical predictions.}}
\label{MCMPI_total}
\end{center}
\end{figure}
In this graph we used for $N=2,3,5,10,25$ the values $(n,m) = (10^2, 5\times 10^3), (10^2, 10^4), (10^2, 2\times 10^4), (80, 3\times 10^4), (80, 5\times 10^4)$ respectively.
To calculate $Z_{AB}$ several sets, labeled by $p$, of random variables $z_A = z_i^{\alpha}$ are generated by a standard Box-Muller transform, the numerical estimation is then used $Z_{AB} \approx \frac{1}{n} \sum_{p =1}^{n} z_{A}^{(p)} z_{B}^{(p)}$ where $p$ is the index that labels the $n$ sets of the Monte Carlo samples, this requires overall generating $nMN$ random variables. In a perfectly calculated average $Z$ would be the identity matrix, but for a finite sample fluctuations around the identity matrix are present. In order to speed up the evaluation we used an approximation by truncating the Taylor series to order $L =25$ in the following form $e^{-2\alpha T \mathbb{V}/M} \approx \mathcal{F} := e^{-4\alpha T/M} \sum_{k=0}^{L} \frac{Q^k}{k!}$, with $\mathbb{V} = 2 {\bf 1}  - Q$ to approximate the exponential in (\ref{mu}) for $N= 10,25$. We kept the full exponential for $N \leq 5$. It was also efficient to split the contributions into diagonal and off-diagonal parts $\sum_{A,B =1}^{MN}Z_{AB} K_{AB} = \sum_{A=1}^{MN} Z_{AA}K_{AA} + 2 \sum_{A<B}^{MN} Z_{AB} K_{AB}$, the off-diagonal terms were further split 
\[
\sum_{A<B} Z_{AB} K_{AB} =  \sum_{\stackrel[]{\alpha}{i<j} } Z_{AB} K_{AB} + \sum_{ \stackrel[]{\alpha < \beta}{ i,j} } Z_{AB} K_{AB}.
\]

The computation time for each contribution grows as the number of components, that is $t_{\rm diag} \sim MN$ and $t_{\rm off;1}\sim M N(N-1)$, $t_{ \rm off;2}\sim N^2 M (M-1)$ for the off-diagonal contributions respectively. An excellent agreement was obtained within the error bars to about 3\% between the numerical simulation of the stochastic dynamics of the brownian chain and the predicted values through the path-integral approach for the average mean square deviation observable $\mu_{N} (T)$. Greater values of $T$ are of course possible at greater computational cost.

\subsection{Calculation from the improved version of MSD}

\subsubsection{Finite $N$ analysis: free and cluster diffusion.}

The short- and long-time regimes are defined by $t\ll t_{0}$ and $t\gg t_{0}$, respectively. In the short-time regime one has $\sigma^{2}_{\ell}\simeq 2D_{0}t$ and $\mu^{2}_{\ell}=\simeq t/3$ for each $\ell$ so that (\ref{MSD-corrections}) has the standard free diffusive behavior, $\sigma^{2}_{N}\left(t\right)\simeq 2D_{0}t$, in the short-time regime. In the long-time regime it will be convenient to split the sum into  $\ell=0$ and $\ell>0$ contributions. For the second contribution it is not difficult to show that $\sigma^{2}_{\ell}\simeq D_{0}/\omega_{\ell}$ and $\mu^{2}_{\ell}\simeq 1/(2\omega_{\ell})$, therefore the modes $\ell>1$ contribute on the MSD just in a constant $C$. The dominant mode in this regime is $\ell=1$ so that the MSD has the following form
\begin{eqnarray}
\sigma^{2}_{N, {\rm imp}}\left(t\right)\simeq \frac{1}{N}\sigma^{2}_{1}\left\{1+\frac{1}{4\gamma}\sigma^{2}_{1}\mu^{2}_{1} V^{(2)}_{11}\right\}+C.
\end{eqnarray}

For $\sigma^{2}_{1}$ and $\mu^{2}_{1}$ we must take into account the regulated zero mode $\omega_{1}=\delta$ in the limit 
$\delta\to 0$. We get for $\sigma^{2}_{1}=2D_{0}t$ and $\mu^{2}_{1}=t/3$. It remains to calculate $\overline{V}^{(2)}_{11}$, in doing so we should be careful since it depends on the orthogonal matrix $\mathcal{O}$ whose elements are given by $\left(\mathcal{O}\right)_{k\ell}=v_{k}^{(\ell)}$ (\ref{eigenvectores}), namely 
\begin{eqnarray}
V^{(2)}_{11}=\sum_{j\ell^{\prime}k}\left(\mathcal{O}^{\dagger}\right)_{1\ell^{\prime}}\left(\mathcal{O}^{\dagger}\right)_{1k^{\prime}}U^{(4)}_{jj\ell^{\prime}k^{\prime}}.
\end{eqnarray}
Notice that $\left(\mathcal{O}^{\dagger}\right)_{\ell k}=v^{(\ell)}_{k}$ and $v^{(1)}_{k}=1/\sqrt{N}$, for each $k$, is the eigenvector corresponding to the zero mode. Using the selection rules found in appendix \ref{apppre22} we find 
\begin{eqnarray}
V^{(2)}_{11}=\frac{1}{N}\sum_{j\ell k}U^{(4)}_{jj\ell k}\simeq 0.
\end{eqnarray}

This means that $\sigma^{2}_{N, {\rm imp}}\left(t\right)=2D_{0}t/N$ for $t\gg t_{0}$.  These results show that for finite $N$ the mean-square displacement $\sigma^{2}_{N, {\rm imp}}\left(t\right)$ (\ref{MSD-corrections}) has the same asymptotic behavior in the short- and large time regimes as the harmonic case (\ref{asymptotic}). Therefore, in both situations the asymptotic behavior of the mean-square displacement, in the short- and in the large time regime, does not depend on the interaction potential.

\subsubsection{Large $N$ analysis: single-file diffusion.}
We will now treat the thermodynamic limit $N\to \infty$. In what follows we will consider only $2\leq \ell\leq N$ since in this limit the contribution from $\ell=1$ becomes negligible. In (\ref{MSD-corrections}) the factor $V^{(2)}_{\ell\ell}$ is  
\begin{eqnarray}
{V}^{(2)}_{\ell\ell}=\sum_{k_{1}, k_{2}, j} v^{(\ell)}_{k_{1}} v^{(\ell)}_{k_{2}}U^{(4)}_{jjk_{1}k_{2}},
\end{eqnarray}
where the orthogonal matrix $\mathcal{O}_{k\ell}=v^{(\ell)}_{k}$ was used. At this point we make use of the selection rules derived in appendix \ref{apppre22}, for this we split each index $k_{1}
$ and $k_{2}$ around $j-1, j, j+1$, and we neglect the irrelevant terms according to the selection rules. Direct calculation produces
\begin{eqnarray}
{V}^{(2)}_{\ell\ell}&=&-\left(\frac{30\kappa_{P}}{a^{2}}\right)\left\{\left[\left(v^{(\ell)}_{1}\right)^{2}+\left(v^{(\ell)}_{N}\right)^{2}\right]\nonumber\right.\\&-&\left.\sum_{j}\left[\left(v^{(\ell)}_{j}-v^{(\ell)}_{j-1}\right)^{2}+\left(v^{(\ell)}_{j+1}-v^{(\ell)}_{j}\right)^{2}\right]\right\}.
\end{eqnarray}

Using now the explicit form of the eigenvectors $v^{(\ell)}_{k}$ (\ref{eigenvectores}), for $2\leq \ell\leq N$ it may be shown that

\begin{eqnarray}
\left(v^{(\ell)}_{1}\right)^{2}=\left(v^{(\ell)}_{N}\right)^{2}=\frac{2}{N}\cos^{2}\left(\frac{\left(\ell-1\right) \pi}{2N}\right),
\end{eqnarray}
and
\begin{eqnarray}
 v^{(\ell)}_{j}-v^{(\ell)}_{j-1}=-\sqrt{\frac{2}{N}}\frac{\sin\left(\frac{\left(\ell-1\right) \pi}{N}\right)\sin\left(\frac{\left(\ell-1\right)\left( j-1\right)\pi}{N}\right)}{\cos\left(\frac{\left(\ell-1\right) \pi}{2N}\right)}.
 \end{eqnarray}
Using these expressions the sum in (\ref{MSD-corrections}) is calculated (see (\ref{sum}) in appendix \ref{B}) in closed form 
\begin{eqnarray}
{V}^{(2)}_{\ell\ell}&=&-\left(\frac{30\kappa_{P}}{a^{2}}\right)\left[\frac{4}{N}\cos^{2}\left(\frac{\left(\ell-1\right) \pi}{2N}\right)-2\frac{\sin^{2}\left(\frac{\left(\ell-1\right)\pi}{N}\right)}{\cos^{2}\left(\frac{\left(\ell-1\right)\pi}{2N}\right)}\right].
\end{eqnarray}

After substituting ${V}^{(2)}_{\ell\ell}$, $\sigma^{2}_{\ell}$, and $\mu^{2}_{\ell }$ in (\ref{MSD-corrections}) we proceed to take the thermodynamic limit $N\to \infty$.  In this limit, is not difficult to observe that the first term in $V^{(2)}_{\ell\ell}$ does not contribute, and hence in this limit the improved MSD $\sigma^{2}_{\rm imp}(t)=\lim_{N\to\infty}\sigma^{2}_{N, {\rm imp}}(t)$ becomes 
\begin{eqnarray}
\sigma^{2}_{\rm imp}\left(t\right)&=&\sigma^{2}(t)+15\left(D_{0}t\rho\right)^{2}\int_{0}^{1}dp~\mathcal{K}\left(4\sin^{2}\left(\frac{\pi p}{2}\right)\frac{t}{t_{0}}\right)+\cdots,
\label{res3}
\end{eqnarray}
where  $\sigma^{2}\left(t\right)$ is the mean-square displacement of the quadratic approximation (\ref{res2}), and $\mathcal{K}\left(x\right):=\left(1-e^{-2x}\right)^2\left[\coth\left(x\right)-x~{\rm csch}^{2}\left(x\right)\right]/x^{2}$. It remains to obtain the asymptotic behavior of the improved mean-square displacement, $\sigma^{2}_{\rm imp}\left(t\right)$, at short, and long times.

We begin with the short-time regime where $t\ll t_{0}$, a series expansion shows that in such case $\mathcal{K}(x)\simeq 8x/3$, and hence the contribution coming from the correction in (\ref{res3}) up to the third order in time can be found indeed using (\ref{series-rep}) expanding up to $t^{3}$ producing
\begin{eqnarray}
\sigma^{2}_{\rm imp}(t)=2D_{0}t-4\frac{D_{0}}{t_{0}}t^2+\left(80\frac{(D_{0}\rho)^{2}}{t_{0}}+8\frac{D_{0}}{t^{2}_{0}}\right)t^{3}+\cdots.
\end{eqnarray}
 Now we turn to the asymptotic behavior in the long-time regime $t\gg t_{0}$. Observe that the function $\mathcal{K}\left(x\right)$ goes to zero for $x\to\infty$, hence the most important contribution of the integral in this regime, occurs when $p$ is near $0$. We can approximate the integral in (\ref{res3}) as $\sim\int_{0}^{\infty}dq~\mathcal{K}\left(\pi^{2} q^{2}\frac{t}{t_{0}}\right)$, therefore the asymptotic behavior obtained in this manner is
  \begin{eqnarray}
 \sigma^{2}_{\rm imp}\left(t\right)\simeq d^{2}_{0} \frac{\theta}{\theta_{0}}\left(\frac{t}{t_{0}}\right)^{\frac{1}{2}}\left[1+\frac{\theta}{\theta_{0}}\left(\frac{t}{t_{0}}\right)
 +\cdots\right],
 \label{MSDimp2}
 \end{eqnarray}
 where $d_{0}=a/\sqrt{(15-10\sqrt{2})\sqrt{2\pi}}$ is a characteristic lenght,  $\theta_{0}=\kappa_{P}a^{2}/(20(3-2\sqrt{2})k_{B})$ is a characteristic temperature, and we recall the characteristic time $t_{0}=\gamma/\kappa_{P}$, with $\kappa_{P}=3\mu_{0}M^{2}\rho^{5}/\pi$ and $\rho=1/a$. 
 \begin{figure}[h!]
\begin{center}
\includegraphics[scale=0.7]{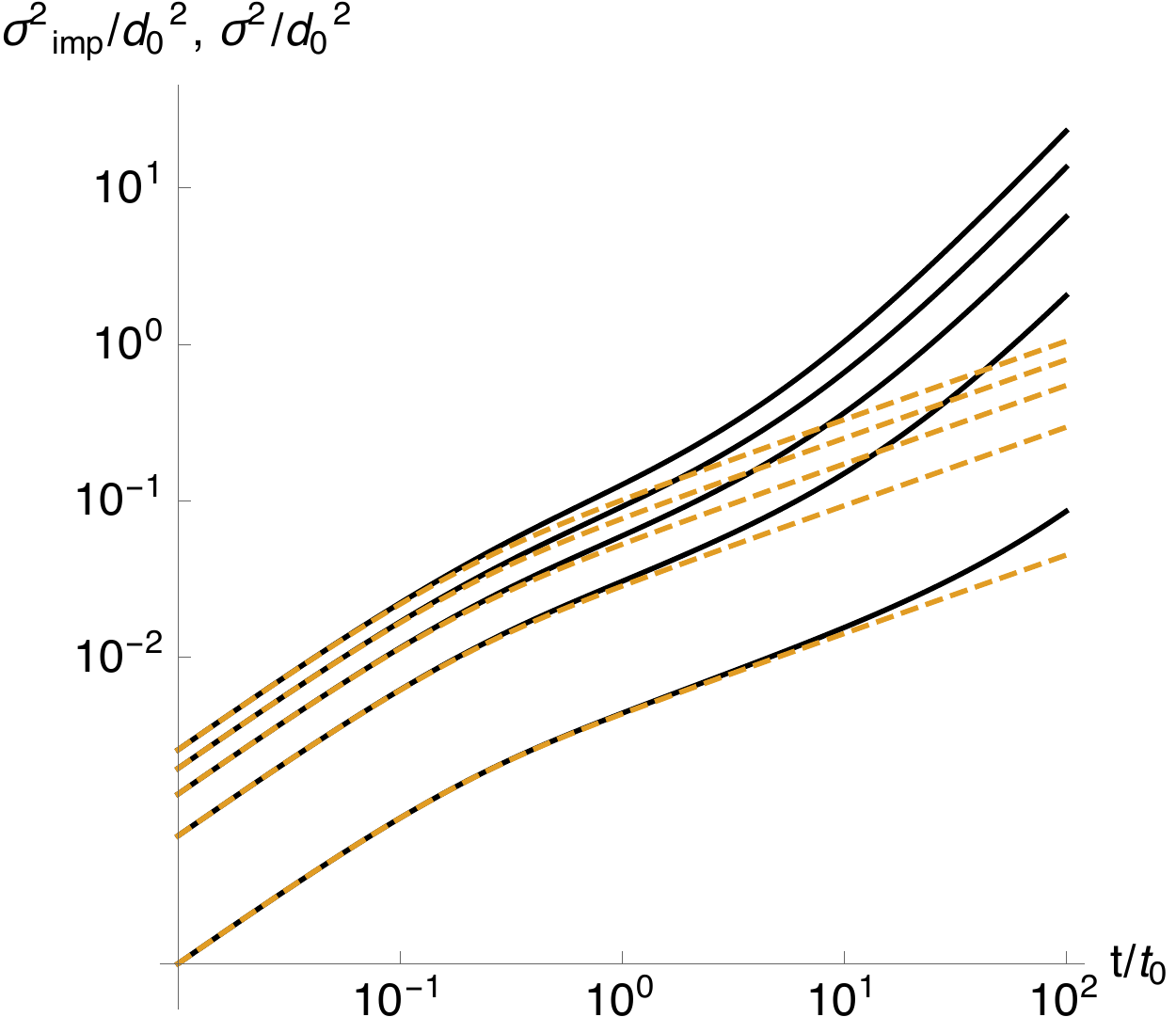}
\caption{{ \small Comparison between improved mean-square displacement, $
\sigma^{2}_{\rm imp}(t)$ (solid line), and $\sigma^{2}(t)$ (dashed line) corresponding to the equations (\ref{res3}), and (\ref{res2}), respectively,  for the values of temperatures $\theta/\theta_{0}=0.009, 0.059, 0.109, 0.159, 0.209$.}}
\label{Imp}
\end{center}
\end{figure}

 In the figure (\ref {Imp}) the behavior of the improved MSD (\ref{res3})  is shown in order to compare against the MSD obtained within the quadratic approximation (\ref{res2}), in it we can appreciate how $\sigma^ {2}_{\rm imp}\left(t\right)$ differs from $\sigma ^ {2}\left(t\right)$ when the temperature $\theta $ increases from $\theta=0.009\,\theta_{0}$ to $\theta=0.209\,\theta_{0}$. These results show that in the long-time regime the behavior of the mean-square displacement differs from the usual behavior of the single-file diffusion $\sim t^{\frac{1}{2}}$. We explain this deviation on the basis that near crystallization of the chain the particles diffuse according to the single-file rule, however when the temperature is increased the SFD rule breaks down due to thermal agitation that produces an increased value of the mean-square displacement. In other words, although the crystalline state displays the SFD behavior in the MSD, our result implies that departing from this state it is easy to break down the SFD behavior. As far as the author know this result has not been reported before in the literature.

\section{Concluding remarks}

In this paper we investigated the Brownian dynamics of a system consisting of interacting particles confined to a straight line through the path integral method. The system was examined near crystallization through the harmonization of the total interaction potential around equidistant positions. Using as a model a harmonic potential the system was analyzed through the many-body Langevin equation in order to compare and validate the result obtained from the path integral formalism. 

The path integral approach we have presented was developed explicitly for a system of interacting particles subjected to a total interaction potential $\Phi\left({\bf x}\right)$ where the configuration vector ${\bf x}\in\mathbb{ R}^{dN}$ being $d$ the dimension of the space, and $N$ the number of particles. Using this formalism we showed that an actual solution of the Smoluchowski equation (\ref{Smoluchowski}) for the joint probability density of finding the system $P\left({\bf x}, {\bf x}_{0}, t\right)$ is given by solving the path integral
\begin{eqnarray}
P\left({\bf x}, {\bf x}_{0}, t\right)=\int_{{\bf x}_{0}}^{{\bf x}}\mathcal{D}{\bf x}\left(\tau\right)
 e^{-\int_{0}^{t}d\tau \left[\frac{\gamma\beta}{4}\left(\dot{\bf x}\left(\tau\right)+\frac{1}{\gamma}\boldsymbol{\nabla}\Phi\left({\bf x}\left(\tau\right)\right)\right)^{2}-\frac{1}{2\gamma}\boldsymbol{\nabla}^2\Phi\left({\bf x}\left(\tau\right)\right)\right]}.\label{SmoluchowskiopathCon}
\end{eqnarray}

This result is the same reported in \cite{Zinn-JustinBook, Chaichian, Olindo2009} in different contexts. In this work we used (\ref{SmoluchowskiopathCon}) to study a system of strongly interacting particles confined to a straight line with $ d=1 $, near crystallization, in particular, the Brownian dynamics of the tracer particle was explored through this formalism using a harmonic interaction potential, $ \Phi\left({\bf x}\right)=\frac {1} {2}\kappa\sum_ {i =1}^{ N-1}\left(x_{i + 1}-x_ {i} -l\right)^2$, finding an excellent agreement between the results implied by the Langevin approach and the path integral at the level of the observable mean square displacement in both analytical and numerical analysis.

Notably we found that when the number of particles $ N $ remains constant, the behavior of the MSD when $ t\ll t_{0}=\gamma / \kappa $, that is at at short times, corresponds to free diffusion, while for long times $ t\gg t_{0} $ the behavior is akin to free diffusion with a reduced diffusion coefficient  $ D_ {\rm eff} = D_{0}/N $; this agrees with the result previously reported in \cite{Delfau2011}. In the thermodynamic limit $ N \to\infty $ it was possible to find a closed formula for the MSD  for all time values in terms of Bessel functions (that is (\ref{res2})). This formula shows that the tracer particle shows a free diffusion behavior in the short-time regime, and displays a typical behavior of single file diffusion $ \sim t^{1/2}$ in the long-time regime.

To investigate the universal character of the $ t^{1/2}$ behavior for confined particles in the line, we treated also the model with a paramagnetic potential,  $ V (r)=\mu_ {0} M^{2}/(4 \pi r ^ {3}) $ near crystallization through the harmonization method \cite{Lizana2010}; here $M$ is the magnetization of the colloidal particles and $\mu_{0}$ the vacuum permeability. Such potential can be thought as having an harmonic form with an effective Hook coefficient given by $ \kappa_ {P} = 3 \mu_ {0} M^2 \rho ^ {5}/\pi $, where $ \rho $ is the density of the system, within a relative error of the order $ 10^{- 2} $. In this approximation the same behavior as in the above harmonic interaction is observed for the mean-square displacement.

In addition an improved MSD was obtained in (\ref{SmoluchowskiopathCon}) considering anharmonic terms in the potential $\Phi({\bf x})$ around the mechanical equilibrium positions,  within the quadratic potential approximation at the Lagrangian level; we also calculated the subdominant term when the temperature is increased off the crystallization state. Such rise in temperature breaks the SFD character $\sim t^{1/2}$ for sufficiently long times, this means that the crystalline Brownian chain also complies with the universal behavior of single file diffusion, while increasing the temperature breaks this $ t^ {1/2}$ behavior, an indication that the diffusion of the tracer particle is enhanced by a higher temperature.

The method developed here can be extended in several directions. For interacting particle systems of dimension $d$, it is possible to analyze the weak interaction regime using the effective potential, $V_{\rm eff}({\bf x})$ as a perturbative term. This procedure constitutes an alternative method \cite{Dhont} to study a diluted system subjected to different external fields (e.g. shear, gravitational or electric field) acting on the particles using an $N-$particle path integral representation.

Following a different direction it is possible to include hydrodynamic interactions by replacing Smoluchowski operator in (\ref{Smoluchowski})  by $\boldsymbol{\nabla}_{a}\mathbb{D}^{ab}\boldsymbol{\nabla}_{b} \cdot+\beta\boldsymbol{\nabla}_{a}\mathbb{D}^{ab}\left[\left(\boldsymbol{\nabla}\Phi\right)_{b}\cdot\right]$, where $\mathbb{D}$ is the  hydrodynamic diffusion tensor; this problem in particular can be reinterpreted as the diffusion of a particle in a higher dimensional Riemannian manifold with a metric tensor proportional to $\mathbb{D}^{-1}$, where a path integral formulation can be also be implemented. The method can be also be used in Monte-Carlo  simulations to model more complex diffusion phenomena.

\ack

P.C.V. acknowledges financial support by CONACyT Grant No. 237425 and Red Temática de la Materia Condensada Blanda, and O.V.R acknowledges financial support from Postdoctoral Grant by  PRODEP (DSA/103.5/16/10437).

\appendix

\section{Deduction of the effective potential $V_{\rm eff}\left({\bf x}\right)$}

Consider a function $U\neq 0$ over $\mathbb{R}^{3N}$ such that  $P/U\equiv \mathbb{P}$  satisfies an equation similar to Schr\"odinger's (with imaginary time). After substituting $P=\mathbb{P}U$ in Smoluchowski equation we get $U\frac{\partial \mathbb{P}}{\partial t}=D_{0}\boldsymbol{\nabla}^2\left(U\mathbb{P}\right)+D_{0}\beta\boldsymbol{\nabla}\cdot\left(U\mathbb{P}\boldsymbol{\nabla}\Phi\right)$.
Now use
\begin{eqnarray}
\boldsymbol{\nabla}^{2}\left(U\mathbb{P}\right)&=&U\boldsymbol{\nabla}^{2}\mathbb{P}+2\boldsymbol{\nabla}\mathbb{P}\cdot\boldsymbol{\nabla}U+\mathbb{P}\boldsymbol{\nabla}^{2}U,\nonumber\\
\boldsymbol{\nabla}\cdot\left(U\mathbb{P}\boldsymbol{\nabla}\Phi\right)&=&\left(\boldsymbol{\nabla}U\cdot \boldsymbol{\nabla}\Phi\right)\mathbb{P}+U\boldsymbol{\nabla}\left(\mathbb{P}\boldsymbol{\nabla}\Phi\right),
\end{eqnarray}
to obtain
\begin{eqnarray}
\frac{\partial \mathbb{P}}{\partial t}&=&D_{0}\boldsymbol{\nabla}^2\mathbb{P}+\left(2D_{0}\frac{\boldsymbol{\nabla}U}{U}+D_{0}\beta\boldsymbol{\nabla}\Phi\right)\cdot\boldsymbol{\nabla}\mathbb{P}\nonumber\\
&+&\left(D_{0}\frac{\boldsymbol{\nabla}^2U}{U}+D_{0}\beta\left(\boldsymbol{\nabla}\log U\right)\cdot\boldsymbol{\nabla}\Phi+D_{0}\beta\boldsymbol{\nabla}^2\Phi\right)\mathbb{P}.
\label{eq1}
\end{eqnarray}
In this manner we choose $U$ in such manner that the coefficient of the first derivative term  $\boldsymbol{\nabla}\mathbb{P}$ vanishes, i.e. $2D_{0}\boldsymbol{\nabla}\log U+D_{0}\beta\boldsymbol{\nabla}\Phi=0$. 
From this equation we find $U({\bf x})=K\left({\bf x}_{0}\right)\exp\left(-\frac{1}{2} \beta\Phi\left({\bf x}\right)\right)$. Notice that $U$ is not the equilibrium solution. The constant $K$ is chosen so that $K=\exp(+\frac{1}{2}\beta\Phi\left({\bf x}_{0}\right))$ holds, and consequently $\mathbb{P}({\bf x},{\bf x}_{0}, t)$ satisfies the initial condition $\mathbb{P}\left({\bf x}, {\bf x}_{0}, 0\right)=\delta\left({\bf x}-{\bf x}_{0}\right)$. 
Now collect the coefficient of $\mathbb{P}$ in equation (\ref{eq1}) to obtain the following Schr\"odinger-type equation 

\begin{equation}
    -\frac{\partial \mathbb{P}}{\partial t}=-D_{0}\boldsymbol{\nabla}^2\mathbb{P}+V_{{\rm eff}}\left({\bf x}\right)\mathbb{P},
\end{equation}
where the resulting ''effective potential'' is $V_{\rm eff}\left({\bf x}\right)$, given by (\ref{effectivepotencial}).

\section{Matrix diagonalization, and useful identities}\label{B}

\subsection{Matriz diagonalization}
In what follows we derive the closed form for the eigenvalues and the characteristic polynomial of the $ N\times N$ matrix

\begin{equation} \label{A1}
\mathbb{V}=\left(\begin{array}{cccccc}
1&-1&0&\cdots&\cdots&0\\
-1&2&-1&\cdots&\cdots&0\\
0&-1&2&-1&\cdots&0\\
\vdots&\vdots&\vdots&\ddots&\ddots&\vdots\\
0&0&\cdots&-1&2&-1\\
0&0&0&\cdots&-1&1
\end{array}\right), 
\label{TridiagonalMatrix}
\end{equation}
to this end let the characteristic polynomial be $p_N (\lambda)  = \det ( \mathbb{V}  - \lambda {\bf 1}) $, an expansion in minors about the first row gives $p_N = (1-\lambda) \delta_{N-1} - \delta_{N-2}$ where $\delta_N$ reffers to the determinant of the $N\times N$ matrix

\[
\delta_N (\lambda) =\left\vert \begin{array}{cccccc}
2-\lambda&-1&0&\cdots&\cdots&0\\
-1&2-\lambda&-1&\cdots&\cdots&0\\
\vdots&\vdots&\vdots&\ddots&\ddots&\vdots\\
0&0&\cdots&-1&2-\lambda&-1\\
0&0&0&\cdots&-1&1-\lambda
\end{array}\right\vert, 
\]
expanding likewise around the last row one gets the relation $\delta_N = (1-\lambda) \Delta_{N-1}  - \Delta_{N-2}$ where now we have introduced the determinant

\[
\Delta_N (\lambda) =\left\vert \begin{array}{cccccc}
2-\lambda&-1&0&\cdots&\cdots&0\\
-1&2-\lambda&-1&\cdots&\cdots&0\\
\vdots&\vdots&\vdots&\ddots&\ddots&\vdots\\
0&0&\cdots&-1&2-\lambda&-1 \\
0&0&\cdots&0&-1&2-\lambda
\end{array}\right\vert, 
\]
therefore we have the relation 
\[
p_N (\lambda) = (1-\lambda)^2 \Delta_{N-2} - 2 (1- \lambda) \Delta_{N-3} + \Delta_{N-4},
\]
and the problem falls back to finding an expression for $\Delta_N (\lambda)$. A minors expansion about the first row, leads to the relation  $\Delta_N = (2-\lambda) \Delta_{N-1} - \Delta_{N-2} $; to solve this recursion we give the Ansatz $\Delta_N =  ar_{+}^N + b r_{-}^{N}$ where $r_{\pm}$ are the roots of the quadratic equation that follows from the Ansatz, namely $r_{\pm} = (2-\lambda \pm \sqrt{(2 -\lambda)^2 - 4} )/2$. At this point we change variable $1- \frac{\lambda}{2} = \cos \varphi$ so that $r_{\pm} = e^{\pm i \varphi}$. Now the constants $a=b=\frac{1}{2}$ are determined from the initial conditions $\Delta_1 = 2\cos \varphi$, $\Delta_2 = 4 \cos^2 \varphi-1 $, resulting into $\Delta_N = \sin[(N+1)\varphi]/\sin \varphi$. We obtain the following closed expression for the characteristic polynomial $p_N  = 2 (\cos \varphi -1) \sin (N \varphi) /\sin \varphi$, that can be expressed as
\[
p_N (\lambda) = -\lambda U_{N-1} \left(\mbox{\small{$1-\frac{\lambda}{2}$}} \right),
\]
where $U_N$ is the $N$-th Chebyshev polynomial of the second kind. The roots of the characteristic polynomial include a zero mode $\lambda=0$ and others that can altogether be interpreted geometrically in terms of $\varphi$ as the real parts of the vertices of an unit regular $n$-gon in the complex plane being  $\varphi = \frac{k-1}{N}\pi$ with $k = 1,2, \cdots, N$. The full spectrum of (\ref{A1}) is found to be
\begin{equation}
\lambda_k  = 4 \sin^2 \left(\frac{\left(k-1 \right)\pi}{2 N} \right), \qquad 1 \leq k \leq N.
\end{equation}

The corresponding eigenvectors of (\ref{TridiagonalMatrix}) can be found using the eigenvalue equations for $k=2,\cdots,  N-1$:
\begin{eqnarray}
v^{(\ell)}_{1}-v^{(\ell)}_{2}&=&\lambda_{\ell}v^{(\ell)}_{1},\nonumber\\
-v^{(\ell)}_{N-1}+v^{(\ell)}_{N}&=&\lambda_{\ell}v^{(\ell)}_{N},\nonumber\\
-v^{(\ell)}_{k-1}+2v^{(\ell)}_{k}-v^{(\ell)}_{k+1}&=&\lambda_{\ell}v_{k}^{(\ell)}.\label{eigenvalueeq}
\end{eqnarray}
 The value of $v^{(\ell)}_{1}=\mathcal{N}^{\ell}$ is fixed by the normalization condition, hence we can find the eigenvectors solving the recurrence relation implied by (\ref{eigenvalueeq}) 
\begin{eqnarray}
v^{(\ell)}_{k+1}+(\lambda_{(\ell)}-2)v^{\ell}_{k}+v^{(\ell)}_{k-1}=0,\label{recrel}
\end{eqnarray}
with the initial conditions $v^{(\ell)}_{2}=(1-\lambda_{\ell})\mathcal{N}^{\ell}$ and $v^{(\ell)}_{3}=\left((2-\lambda_{\ell})(1-\lambda_{\ell})-1\right)\mathcal{N}^{\ell}$ for (\ref{recrel}). For each value of $\ell$ (\ref{recrel}) is a linear recurrence relation with well posed initial conditions. The solution of the recurrence relation is obtained by the standard method proposing $v^{(\ell)}_{k}=c_{+}r_{+}^{k-1}+c_{-}r_{-}^{k-1}$, where $c_{\pm}$ is to be determined from the initial conditions.

A straightforward calculation of the eigenvectors reveals that the eigenvector corresponding to the zero eigenvalue is $v^{(1)}_{k}=1/\sqrt{N}$ whereas the remaining eigenvectors are
\begin{eqnarray}
v^{(\ell)}_{k}=
\frac{1}{\sqrt{2N}}\frac{\cos\left(\frac{(\ell-1)k \pi}{N}\right)+\cos\left(\frac{(\ell-1)(k-1)\pi}{N}\right)}{\cos\left(\frac{(\ell-1)\pi}{2N}\right)}. \label{eigenvectores}
\end{eqnarray}
The components of the  orthogonal matrix $\mathcal{O}$ are given by $\left(\mathcal{O}\right)_{k \ell}=v^{(\ell)}_{k}$.




\subsection{Useful identities}

The following identities are useful for the calculation of the MSD
\begin{eqnarray}
I_0'(x)&=& I_1 (x),\nonumber\\
 I'_1 (x) &=& I_0 (x) - \frac{I_1 (x)}{x},\label{id2}
 \end{eqnarray}
 where $I_{0}(x)$ and $I_{1}(1)$ are  the modified Bessel functions of the first kind of order $0$ and $1$,  respectively. We also used the identities 
\begin{eqnarray}
\int_{0}^{\frac{\pi}{2}}d\theta~\sin^{(2k-2)}\theta=\frac{\sqrt{\pi}}{2}\frac{\Gamma\left(k-\frac{1}{2}\right)}{\Gamma\left(k\right)},
\label{id}
\end{eqnarray} 
and
  \begin{eqnarray}
 \sum_{j=1}^{N}\sin^{2}\left({\frac{\left(\ell-1\right)\left(j-1\right)\pi}{N}}\right)= \frac{N}{2}.\label{sum}
 \end{eqnarray}

\section{Estimation of numerical error}

There are three sources of error present in our numerical calculation that give rise to error bars in figure (\ref{MCMPI_total}), these are: i. The quadrature error $\epsilon_M$ for approximating the integral as a Riemann sum of $M$ terms ii. The truncation error $\epsilon_L$ in the exponential series at order $L$
iii. The error due to the finite size $n$ of the statistical Monte Carlo sample $\epsilon_n$. To estimate each error consider all three sources of error

\begin{eqnarray} \nonumber
\Delta \mu(T)& =& \frac{2T}{N} \left[ \int_0^1{\rm tr} \left( e^{-2\alpha T \mathbb{V}/M} \right)-  \sum_{\alpha=1}^M \frac{1}{M} {\rm tr} \left( e^{-2\alpha T \mathbb{V}/M} \right) \right]  \\
& +& \frac{2T}{N} \sum_{\alpha =1}^M \frac{1}{M} {\rm tr} \left( e^{-2\alpha T \mathbb{V}/M} - \mathcal{F} \right) - \frac{2T}{N} \sum_{\alpha =1}^M \frac{1}{M} {\rm tr} (\mathcal{F} \Phi) \nonumber \\
&=& \epsilon_M + \epsilon_L + \epsilon_n,
\end{eqnarray}
where we have introduced the matrix $\Phi = 1 - Z$ and used $\mathcal{F} = e^{-4\alpha T/M} \sum_{k=0}^L \frac{Q^k}{k!}$.

Using the Euler-Maclaurin formula we may show that the truncation by $M$ underestimates the integral by the amount 

\[
\epsilon_M =\frac{T}{MN} \left(\sum_{\ell =0}^{N-1}e^{-2T\lambda_\ell} - N \right) + O(1/M^2), 
\]
a good estimation of this error is
\begin{equation}
\epsilon_M \approx - \frac{T}{M}.
\end{equation}
This is the dominant source of error. A closed form for the error $\epsilon_L$ can be found based upon the exact partial sum of the exponential series

\begin{equation}
\epsilon_L = \frac{2T}{MN L!}\sum_{\alpha =1}^{M}\sum_{\ell =0}^{N-1} e^{-2\alpha T \lambda_\ell /M} \gamma (L+1, 4\alpha T \cos (\pi \ell /N)/M).
\end{equation}
Where $\gamma$ is the lower incomplete gamma function. This second source of error turns out to play no role in our calculations since for $N\leq 5$ it vanishes and for $N=10,25$ is always $\epsilon_L < 2\times 10^{-5}$, at least three orders of magnitude smaller than $\epsilon_M$.

Finally the error due to statistical fluctuations on a finite size $n$ sample is 
\[
\epsilon_n = - \frac{2T}{MN}\sum_{\alpha = 1}^M \sum_{k=1}^L e^{-4\alpha T/M} \frac{{\rm tr} (Q^k \Phi)}{k!}.
\]
Let the eigenvalues of $Q$ be $q_\ell$, consider the inequality $Tr (Q^k \Phi) \leq \sqrt{\sum_{\ell} q_\ell^{2k} \sum_{r,s} \Phi_{rs}^2} $ and the estimate of the elements of $\Phi$ given by the central limit theorem $\Phi_{r,s} \sim \pm (1- \delta_{rs})/\sqrt{n} $.

A reasonable estimate of the fluctuation error for finite $L$ is then found to be

\begin{equation}
\epsilon_n \approx \frac{2T}{M\sqrt{n}}\sum_{\alpha=1}^M \sum_{k=1}^{L} e^{-4\alpha T /M} \frac{2^k}{k!}\left(\sum_{\ell =0}^{N-1} \cos^{2k} (\pi \ell /N)\right)^{1/2}.
\end{equation}
taking $L \to \infty$ in our data does not change the error estimate substantially so $L=25$ is used throughout all estimations.

\section{Selection rules from the anaharmonic terms}\label{apppre22}

The third  coefficient is given by
\begin{eqnarray}
U^{(3)}_{\ell k m}=\frac{5\kappa_{P}}{a}\left\{\begin{array}{cc}
\zeta^{(6)}_{k}-\zeta^{(6)}_{N-k+1}, &~~{\rm for}~ \ell=m=k\\
\\
-\frac{\left(\ell-k\right)}{\left|\ell-k\right|^7}, &~~{\rm for}~ m=k\\
\\
0, & {\rm for}~~  {\rm otherwise}.
\end{array}
\right.
\end{eqnarray}
By numerical inspection we observe that the central terms satisfy $U^{(3)}_{111}=U^{(3)}_{NNN}=\frac{5\kappa_{P}}{a}\left(1+\epsilon_{111}\right)$, $U^{(3)}_{kkk}=\frac{5\kappa_{P}}{a} \epsilon_{kkk}$, where both $\epsilon_{111}$ and $\epsilon_{kkk}$ are less than $10^{-2}$, moreover for the particles in the bulk the later terms satisfy $\epsilon_{kkk}\leq 10^{-9}$. The off-central terms satisfy $U^{(3)}_{k+1, kk}=-U^{(3)}_{k-1, kk}=-\frac{5\kappa_{P}}{a}$  while $U^{(3)}_{k\pm\ell, kk}=\mp \frac{5\kappa_{P}}{a}\epsilon_{\ell kk}$, where $
\epsilon_{\ell k k}<1.6\times 10^{-2}$ for $\ell\geq 2$. The fourth coefficient is given by
\begin{eqnarray}
U^{(4)}_{\ell k m n}=\frac{30\kappa_{P}}{a^2}\left\{\begin{array}{cc}
H_{k-1}^{(7)}+H_{N-k}^{(7)}, & {\rm for}~ \ell=k=m=n \\
\\
-\frac{1}{\left|\ell-k\right|^{7}}, & {\rm for}~ \ell=m=n\neq k\\
\\
\frac{1}{\left|\ell-k\right|^{7}}, & {\rm for}~ \ell=m\neq n= k\\
\\
0 & {\rm otherwise}
\end{array}\right. 
\end{eqnarray}

By numerical inspection the central terms satisfy $U^{(4)}_{1111}=U^{(4)}_{NNNN}=\frac{30\kappa}{a^2}\left(1+\epsilon_{1111}\right)$,  $U^{(4)}_{kkkk}=\frac{30\kappa}{a^2}\left(2+\epsilon_{kkkk}\right)$ where both $\epsilon_{1111}$ and $\epsilon_{kkkk}$ are of the order of $10^{-3}$. Finally, the off-central terms satisfy $U^{(4)}_{kkk,k\pm 1}=-U^{(4)}_{k,k\pm1,k, k\pm 1}=-\frac{30\kappa}{a^2}$ and $U^{(4)}_{kkk, k\pm \ell}=-U^{(4)}_{kk, k\pm \ell, k, k\pm \ell}\approx -\frac{30\kappa}{a^{4}}\epsilon_{kkk \ell}
$ where $\epsilon_{kkk\ell}<7.9\times 10^{-3}$ for $\ell\geq 2$. 

\section{Partition function and the Green function
.}\label{applast} To calculate $\left<\overline{V}_{I}\left({\bf z}_{\rm cl}(\tau)+{\bf q}(\tau)\right)\right>$, we use (\ref{defaverage}). 
To achieve this we perform the coordinate transformation ${\bf q}=\mathcal{O}{\bf r}$, where $\mathcal{O}$ is the orthogonal matrix that diagonalizes $\mathbb{U}$. In this manner introducing a source term ${\bf J}$ in the partition function $\mathbb{Z}\left[{\bf J}(\tau)\right]$ we define 
\begin{eqnarray}
\mathbb{Z}\left[{\bf J}\left(\tau\right)\right]=\prod_{\ell=1}^{N}\oint \mathcal{D}{r}_{\ell}(\tau)~
e^{-\int_{0}^{t}d\tau\left(\mathcal{L}_{0, \ell}\left(\dot{ r}_{\ell}(\tau), { r}_{\ell}(\tau)\right)- { J}^{\mathcal{O}}_{\ell}(\tau){ r}_{\ell}(\tau)\right)},
\label{partitionfunction2}
\end{eqnarray}
where ${\bf J}^{\mathcal{O}}=\mathcal{O}^{\dagger}{\bf J}$, and the Lagrangian $\mathcal{L}_{0, \ell}\left(\dot{ r}_{\ell}(\tau), { r}_{\ell}(\tau)\right)$ is given by (\ref{Lagrangian0}). We recall the boundary conditions $r_{\ell}(0)
=r_{\ell}(t)=0$, and follow the standard procedure to complete the square in the argument of the exponential, this leads to
\begin{eqnarray}
\mathbb{Z}\left[{\bf J}\left(\tau\right)\right]=\mathbb{Z}\left[0\right]\prod_{\ell=1}^{N}
e^{\frac{1}{2}\int_{0}^{t}d\tau \int_{0}^{t}d\tau^{\prime}J^{O}_{\ell}(\tau)G_{\ell}(\tau, \tau^{\prime})J^{O}_{\ell}(\tau^{\prime})},
\label{partitionfunction3}
\end{eqnarray}
where $G_{\ell}\left(\tau, \tau^{\prime}\right)$ is the Green function associated with the Helmholtz-like operator 
$
-\mathcal{D}^{2}:=-\frac{1}{2D_{0}}\left(\frac{\partial}{\partial\tau}+\omega_{\ell}\right)^{2}
$ 
with Dirichlet boundary conditions $G(0, \tau^{\prime})=G(t, \tau^{\prime})=0$. 

Using the standard methods one could solve for the Green function, however this is not necessary for the present work.
The factor $\mathbb{Z}\left[0\right]$ can be found explicitly \cite{Zinn-JustinBook}, but it plays no role whatsoever for our purposes.

The potential $V_{I}({\bf x})$ after the coordinate transformation ${\bf y}=\mathcal{O}{\bf z}$ can be obtained by replacing $y\to z$ and $\mathbb{V}^{(n)}\to \overline{\mathbb{V}}^{(n)}$, where 
$\overline{\mathbb{V}}^{(n)}=\mathcal{O}^{\dagger}\otimes\mathcal{O}^{\dagger} \cdots \otimes \mathcal{O}^{\dagger}\mathbb{V}^{(n)}$  are the transformed tensors, after the coordinate transformation. After the substitution ${\bf z}\to {\bf z}_{\rm cl}+{\bf r}$ we take the expectation value using (\ref{partitionfunction3}). Notice that  expectation values of an odd number of $r^{\ell}\left(\tau\right)$ are zero when ${\bf J}=0$.  

Therefore the only terms that contribute in the expectation value of $V_{I}\left({\bf z}_{\rm cl}+{\bf r}\right)$ are
\begin{eqnarray}
\left<V_{I}\left({\bf z}_{\rm cl}+{\bf Q}\right)\right>&\simeq&\overline{V}^{(1)}_{k}z^{k}_{\rm cl}(\tau)+\overline{V}^{(2)}_{k\ell}z^{k}_{\rm cl}(\tau)z^{\ell}_{\rm cl}(\tau)+\overline{V}^{(2)}_{k\ell}\left<r^{k}(\tau)r^{\ell}(\tau)\right>,
\end{eqnarray}
where the correlation function is $\left<r^{k}(\tau)r^{\ell}(\tau)\right>=\delta^{k\ell}G_{\ell}\left(\tau, \tau\right)$, and ${z}^{\ell}_{\rm cl}
$ is the classical solution (\ref{classicalsol}).  Note that the coefficient $\mathbb{V}^{(2)}$ already has information of the quartic term in the Taylor expansion (\ref{TaylorExpansion}). 

After calculating the integrals  $I_{1}:=\int_{0}^{t}d\tau z^{k}_{\rm cl}(\tau)$,  $I_{2}:=\int_{0}^{t}d\tau G_{\ell}(\tau, \tau)$, and $I_{3}:=\int_{0}^{t}d\tau z^{k}_{\rm cl}(\tau)z^{\ell}_{\rm cl}(\tau)$ results
\begin{eqnarray}
I_{1}=\frac{1}{\omega_{k}}\left(z^{k}+z^{k}_{0}\right)\tanh\left(\frac{\omega_{k}t}{2}\right),
\end{eqnarray}
and
\begin{eqnarray}
I_{3}&=&\frac{1}{\omega^{2}_{k}-\omega^{2}_{\ell}}\left[\left(z^{k}z^{\ell}-z^{k}_{0}z^{\ell}_{0}\right)\left(\omega_{k}\coth(\omega_{k}t)-\omega_{\ell}\coth(\omega_{\ell}t)\right)\right.\nonumber\\
&-&\left.\left(z^{k}z^{\ell}_{0}+z^{\ell}z^{k}_{0}\right)\left(\omega_{k}{\rm csch}(\omega_{k}t)-\omega_{\ell}{\rm csch}(\omega_{\ell}t)\right)\right].
\end{eqnarray}

We point out that the function $\mathcal{P}({\bf x}, {\bf x}_{0}, t)$ can be written approximately as 
\begin{eqnarray}
\mathcal{P}({\bf x}, {\bf x}^{\prime}, t)\simeq \mathbb{Z}\left[0\right]
e^{-\int_{0}^{t}d\tau\left<V_{I}\left({\bf z}_{\rm cl}(\tau)+r_{\ell}(\tau)\right)\right>}.
\label{Pscript}
\end{eqnarray}
and since $I_{2}$ depends only on time $t$ this term does not play any role, and neither does $\mathbb{Z}[0]$ as it cancels out when the normalization constant in (\ref{PDFnonharmonic}) $\mathcal{Z}$ is considered.


\section*{References}
\bibliographystyle{ieeetr}
\bibliography{BrownianChain.bib}

\end{document}